\documentclass[usenatbib]{mn2e}

\usepackage{psfig,morefloats,url}
\usepackage{graphicx}
\usepackage{natbib}
\usepackage{amssymb}

\newcommand{\psr}{PSR~J1738$+$0333}

\title[The most stringent test of scalar-tensor gravity.]{
The~relativistic~pulsar-white~dwarf~binary~PSR~J1738+0333
II. The most stringent test of scalar-tensor gravity} 

\author[Freire, Wex, Esposito-Far\`ese, Verbiest et al.]
{Paulo~C.~C.~Freire$^{1}$\thanks{E-mail: pfreire@mpifr-bonn.mpg.de},
Norbert~Wex$^{1}$,
Gilles Esposito-Far\`ese$^{2}$,
Joris~P.~W.~Verbiest$^{1}$,  
\newauthor
Matthew~Bailes$^{3}$,
Bryan~A.~Jacoby$^{4}$,
Michael Kramer$^{1,5}$,
Ingrid~H.~Stairs$^{6}$,
\newauthor
John~Antoniadis$^{1}$\thanks{Member of the International Max Planck Research School (IMPRS) for Astronomy and Astrophysics at the Universities of Bonn and Cologne.},
and Gemma~H.~Janssen$^{5}$
\\
$^{1}$ Max-Planck-Institut f\"ur Radioastronomie, Auf dem H\"ugel 69,
D-53121 Bonn, Germany\\
$^{2}$ UPMC-CNRS, UMR7095, Institut d'Astrophysique de Paris, GReCO, 98bis boulevard Arago, F-75014 Paris, France\\
$^{3}$ Centre for Astrophysics and Supercomputing, Swinburne
University of Technology, P.O. Box 218, Hawthorn, VIC 31122, Australia\\
$^{4}$ Affiliated with The Aerospace Corporation, 15049 Conference Center Drive, Chantilly, VA 20151-3824, USA\\
$^{5}$ University of Manchester, Jodrell Bank Centre for Astrophysics, Alan Turing Building, Manchester M13 9PL, United Kingdom\\
$^{6}$ Department of Physics and Astronomy, University of British Columbia, 6224 Agricultural Rd., Vancouver, BC V6T 1Z1, Canada\\
}
\begin{document}

\maketitle

\begin{abstract} 
We report the results of a 10-year timing campaign on \psr, a 5.85-ms
pulsar in a low-eccentricity 8.5-hour orbit with a low-mass white dwarf 
companion. We obtained 
17376 pulse times of arrival with a stated uncertainty smaller than $5\,\mu$s 
and weighted residual rms of $1.56\,\mu$s. The large number and precision of 
these measurements allow highly significant estimates of the
 proper motion $\mu_{\alpha,\delta} = (+7.037\,\pm\,0.005, +5.073 \,\pm\, 
0.012)$\,mas\,yr$^{-1}$, parallax $\pi_x \,=\, (0.68\,\pm\,0.05)$\,mas and a 
measurement of the apparent orbital decay, $\dot{P}_b \,=\, (-17.0\,\pm\, 3.1) 
\times 10^{-15}\,{\rm s\,s^{-1}}$ (all 1-$\sigma$ uncertainties). The   
measurements of $\mu_{\alpha,\delta}$ and $\pi_x$ allow for
a precise subtraction of the kinematic contribution to the observed orbital decay;
this results in a significant measurement of the intrinsic orbital decay:  
$\dot{P}_b^{\rm Int} \,=\, (-25.9\,\pm\,3.2) \times 10^{-15}\,{\rm  s\,s^{-1}}$. 
This is consistent with the orbital decay from the emission of gravitational waves 
predicted by general relativity, $\dot{P}_b^{\rm GR}\,=\,-27.7^{+1.5}_{-1.9} 
\times 10^{-15}\,{\rm s\,s^{-1}}$, i.e., general relativity passes the test 
represented by the orbital decay of this system. This agreement introduces a 
tight upper limit on dipolar gravitational wave emission, a prediction of 
most alternative theories of gravity for asymmetric binary systems such as this. 
We use this limit to derive the most stringent
constraints ever on a wide class of gravity theories, where gravity involves a
scalar field contribution. When considering general scalar-tensor theories of
gravity, our new bounds are more stringent than the best current solar-system
limits over most of the parameter space, and constrain  the matter-scalar
coupling constant $\alpha_0^2$ to be below the $10^{-5}$ level.
For the special case of the Jordan-Fierz-Brans-Dicke, we obtain the one-sigma
bound $\alpha_0^2 < 2 \times 10^{-5}$, which is within a factor two of the Cassini limit. We 
also use our limit on dipolar gravitational wave emission to constrain a wide  
class of theories of gravity which are based on a generalization of Bekenstein's 
Tensor-Vector-Scalar gravity (TeVeS), a relativistic formulation of Modified 
Newtonian Dynamics (MOND).
\end{abstract}


%%%%%%%%%%%%%%%%%%%%%%%%%%%%%%%%%%%%%%%%%%%%%%%%%%%%%%%%%%%%%%%%%%%%%%%%%%%%%%%%

\section{Introduction}
\label{sec:intro}

Pulsar J1738+0333 is one of seven recycled pulsars discovered in a high-Galactic 
latitude ($15^\circ < |b| < 30^\circ$, $-100^\circ < l < 50^\circ$ Parkes 
multi-beam pulsar survey \citep{jac05,jbo+07}. It has a spin period $P$ of 5.85 
ms and it is located in a low-eccentricity ($e < 4 \times 10^{-7}$) binary 
system with an orbital period ($P_b$) of 8.5 hours; the projected semi-major 
axis of the pulsar orbit ($x$) is 0.3434 light seconds. The Parkes timing 
determined a phase-coherent timing solution with precise spin, orbital and 
astrometric parameters that allowed the detection of the companion of the pulsar 
at optical wavelengths \citep{jac05}. 

The results of the optical observations of the companion are described in detail 
by \citet[henceforth Paper I]{akk+12}; it is a white dwarf (WD), and from its 
spectrum they estimate its mass $M_c\,=\,0.181^{+0.008}_{-0.007}\,\rm M_{\odot}$
(1-$\sigma$, where $\rm M_{\odot}$ represents one Solar mass; we also
define $m_c \equiv M_c/\rm M_{\odot}$). The paper also presents measurements of 
the Doppler shifts of the WD spectral lines which are used to estimate the 
systemic radial velocity $\gamma = (-42 \pm 16)\,\rm km\,s^{-1}$ and the 
semi-amplitude of the WD orbital velocity projected along the line of sight 
$K_c = (171 \pm 5)\, \rm km \, s^{-1}$, consistent (but significantly more 
precise) with the value obtained by \cite{jac05}, $K_c = (181 \pm 27)\, \rm km 
\, s^{-1}$. This measurement is combined with its pulsar equivalent
$K_p = 2 \pi x c / P_b = 21.10336\, \rm km \, s^{-1}$ to estimate
the mass ratio: $q\,\equiv\,M_p/M_c\, = K_c/K_p =\,8.1\,\pm\,0.2$.
Given $M_c$, this implies a pulsar mass
$M_p\,=\,1.46^{+0.06}_{-0.05}\,\rm M_{\odot}$. Finally, from Kepler's
laws they derive an orbital inclination $i = 32\fdg6 \pm
1\fdg0$.

The position pf \psr\, makes it detectable with the Arecibo
Observatory's 305-m William E. Gordon radio telescope, which provides
about 15 times the
sensitivity of the Parkes telescope and therefore allows much more
precise timing. Regular Arecibo observations of \psr\ started in
2003. Given the mass measurements obtained from the optical/radio data,
and the knowledge that $e < 4 \times 10^{-7}$,
general relativity (GR) predicts an orbital decay of
$\dot{P}_b^{\rm GR} = (-27.7^{+1.5}_{-1.9}) \times
10^{-15} \rm s \, s^{-1}$ (henceforth fs\,s$^{-1}$) due
to the emission of gravitational waves (GW).
Our early simulations suggested that using Arecibo we would be
able to measure this effect within a few years.

The organization of the remainder of this paper is as follows: in 
Section~\ref{sec:observations} we describe in detail the observations and how 
the pulse times of arrival and the timing solution were derived. In 
Section~\ref{sec:results} we discuss the detection of the orbital decay of the 
system. We compare it with the prediction from GR, and derive an upper limit for 
any excess GW emission. In Section~\ref{sec:generic}, we use this limit to 
derive an upper limit for the emission of dipolar GWs. In the following 
sections, we discuss the implications of this result for alternative theories of 
gravity, specifically in Section~\ref{sec:STe} for tensor-scalar theories of 
gravity and in Section~\ref{sec:TeVeS} for a class of tensor-vector-scalar 
theories of gravity similar to the theory proposed by \cite{bek04}, which is a 
relativistic formulation of the Modified Newtonian Dynamics (MOND) proposed by 
\cite{mil83}. Finally, in Section~\ref{sec:conclusions}, we summarize our 
results and highlight the prospects opened by continued timing of this system.

%%%%%%%%%%%%%%%%%%%%%%%%%%%%%%%%%%%%%%%%%%%%%%%%%%%%%%%%%%%%%%%%%%%%%%%%%%%%%%%%

\section{Radio timing observations}
\label{sec:observations}

\begin{figure}
  \includegraphics[width=8.5cm]{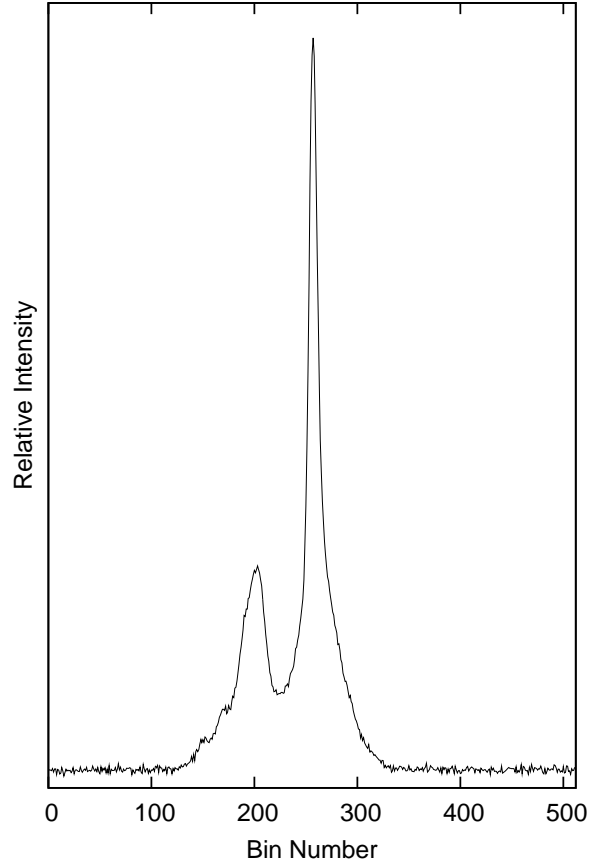}
  \caption{PSR~J1738+0333 average pulse profile at 1400~MHz taken with
    the WAPP spectrometers. A full 512-bin rotation cycle is displayed.
    \label{fig:profile}}
\end{figure}

\subsection{Observational setup and data reduction}

\psr\ was discovered with the 64-m Parkes telescope in an observation
taken in 2001. We started timing it regularly on 2001 September 26
using the 21-cm multi-beam receiver system and the
$2 \times 512 \times 0.5 \,\rm MHz$ filterbank. The Parkes datataking
and processing are described in detail in \cite{jac05}.

In August 2003 we started regular observations of this
MSP using the L-wide receiver of the 305-m William E. Gordon Telescope
at the Arecibo Observatory and the Wide-band Arecibo
Pulsar Processors (WAPPs) to process the signal. The WAPPs are
general-purpose auto-correlators capable of processing a band of up to
100 MHz each \citep{dsh00}.
First, they digitise the incoming voltages as 3-level values. The
machine then calculates the auto-correlation function (ACF) for a
specified number of time lags, in our case 256. These are integrated
for a time $t_{\rm S}$ of 64$\,\mu$s, after which they are written
to disk as 16-bit
integers for later off-line analysis. The four WAPPs were centered at
frequencies $F$ of 1170, 1310, 1410 and 1510 MHz.

The ACFs from the WAPPs are, later on, read from disk during
off-line processing. First, they are Fourier transformed using
Duncan Lorimer's SIGPROC routine ``filterbank'', producing 100
MHz-wide, 256-channel power spectra. To eliminate signal
distortions caused by the edges of the WAPP bands we set the
values of the lower and higher 16 channels to zero, all other
channel values are unchanged. These spectra are then divided into
four sub-bands. Each of these is dedispersed at the DM of the pulsar.
The resulting time series is then folded modulo the pulsar rotational
period, using a polynomial expansion of the pulsar spin phase
as a function of time predicted by the best existing
ephemeris. For each of these sub-bands, the pulses are integrated for
4 minutes as single 512-bin pulse profiles. Only on 
days when the pulsar is consistently faint (the flux density
changes because of 
scintillation caused by the interstellar medium) do we integrate the pulse 
profiles for 600 seconds. All of this processing is done using the
``sig\_foldpow'' routine, written by one of us (IHS).
Given the pulsar's dispersion measure (DM, a measurement of the column
density of free electrons between the Earth and the pulsar)
of 33.77\,cm$^{-3}$\,pc, the dispersive smearing per channel
(with bandwidth $\delta F = 100 \rm MHz / 256 = 0.390625\, MHz$) was
$t_{\rm DM}\,=\, 8.3 \mu {\rm s}\, ({\rm DM / cm^{-3} pc}) (\delta F/ {\rm MHz})
(F / {\rm GHz})^{-3} = $ 68.3, 48.7, 39.0 and 31.8\,$\mu$s per channel
respectively. The total time resolution
$dt = \sqrt{t_{\rm S}^2 + t_{\rm DM}^2}$ was therefore 93.6, 80.4, 75.0
and $71.5\,\mu$s respectively.

Producing a large number of such profiles has many advantages:
a) The separation in frequency allows the use
of polynomial coefficients derived specifically
for the radio frequency of each sub-band in such a way that the
orbital phase of the binary is constant for the pulse arrival time at each
frequency, not constant for a particular time. This is very important
for systems with short orbital periods like \psr, where the
difference in arrival times at the lower and upper edges of the band
caused by dispersion by the ionized interstellar medium corresponds to a
non-negligible shift in the orbital phase.
b) The production of many TOAs in time preserves the
important orbital information contained in them. c) By having
small bandwidths and integration times, we minimize profile
smearing due to imperfections in the ephemeris and DM model. Finally, d) we
take full advantage of the power of scintillation -- in some
occasions, the signal-to-noise ratio in a 25 MHz$\,\times\,$4-minute
subsection is larger than for the whole observation.

%==============================================================================%

%----------

\begin{figure*}
  \includegraphics[width=17.5cm]{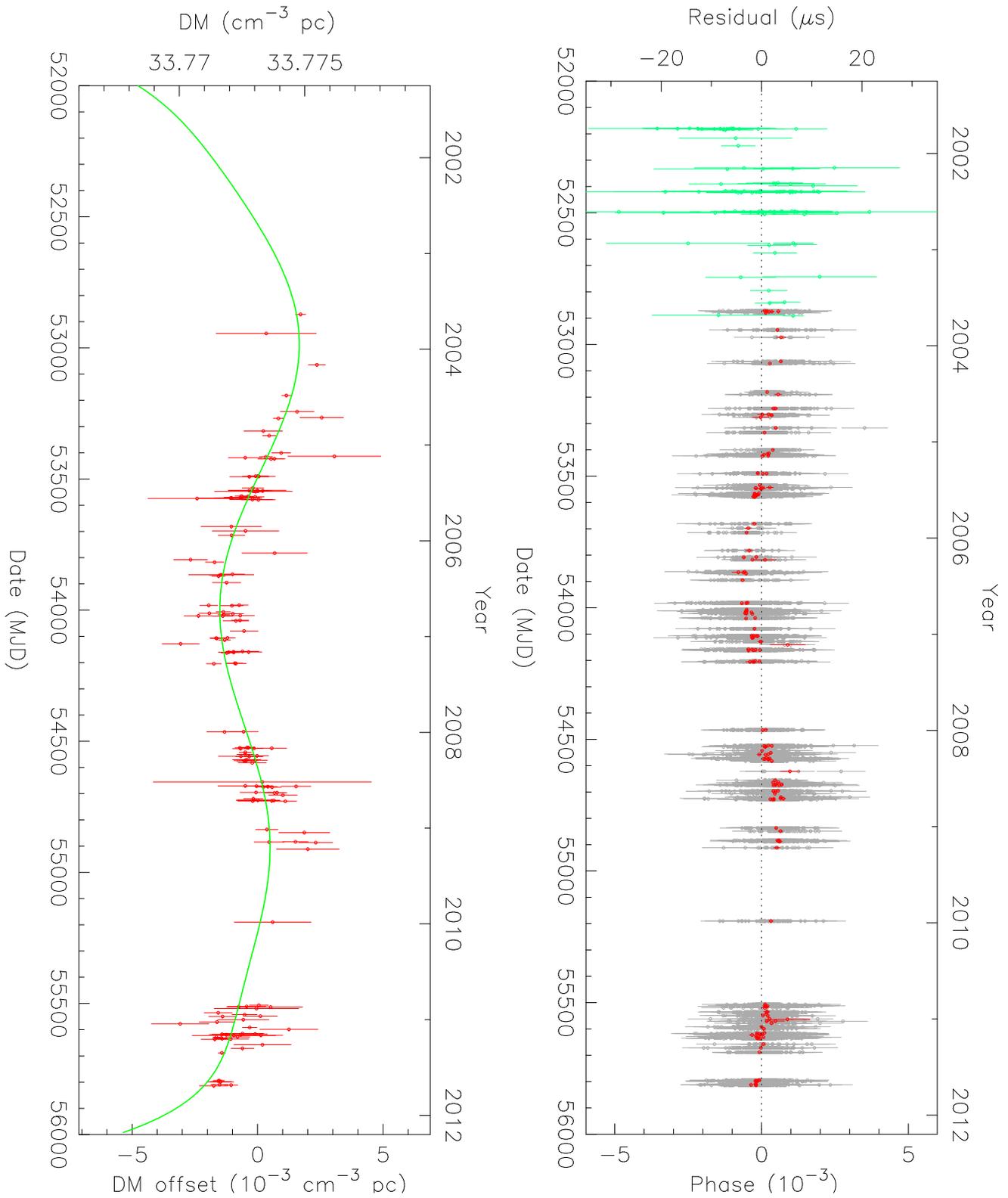}
  \caption{\label{fig:residuals}
    {\em Left:} Daily DM measurements versus time. Given the
    non-linear variation of the DM, it is clear that it must be
    modeled using a high-order polynomial. We display in green the
    eighth-degree polynomial in Table~\ref{tab:parameters}; this fits the
    DM evolution well inside the range where there are DM measurements, but
    diverges fast outside that range.
    {\em Right}: Post-fit residuals versus time. The residuals of the
    Parkes (green) and Arecibo (gray) TOAs were obtained with a
    preliminary ephemeris that assumes a constant DM.
    The averages of the residuals for each sidereal day are indicated in red.
    Notice their similarity with the daily DM averages; this implies that
    implying the former are (mostly) caused by the latter.}
\end{figure*}

%----------

\begin{figure*}
  \includegraphics[width=17.5cm]{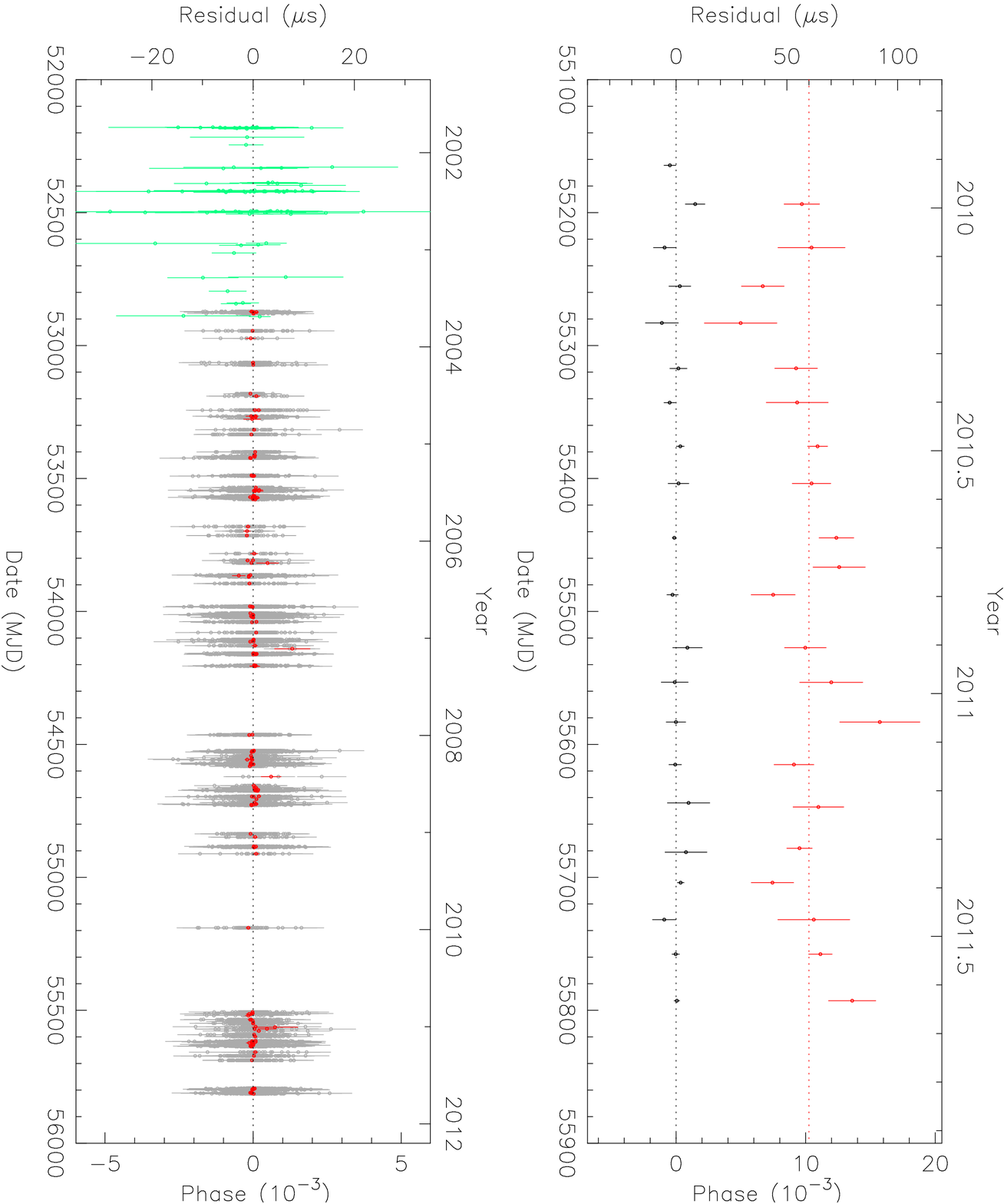}
  \caption{
    {\em Left}: Post-fit residuals versus time. The residuals of the
    Parkes (green) and Arecibo (gray) TOAs were obtained with the timing model 
    presented in Table~\ref{tab:parameters}. 
    The averages of the residuals for 
    each sidereal day are indicated in red.
    {\em Right:} Pre-fit residuals versus time obtained from the
    Westerbork Synthesis Radio Telescope TOAs at 
    1380 MHz (black) and 345 MHz (red). The latter residuals are displayed
    with an offset of 60\,$\mu$s for clarity. These were obtained
    using the timing model presented in Table~\ref{tab:parameters}. 
    The lack of trends in the Westerbork residuals implies
    that our timing solution describes these TOAs correctly.
    These data were not used in the derivation of the timing solution presented
    in this paper; their inclusion does not change any parameter significantly.
    \label{fig:residuals_corrected} }
\end{figure*}

%----------

\subsection{Derivation of times of arrival}
\label{sec:TOAs}

The best pulse profile, resulting from more than 1 
hour of data, is displayed in Fig.~\ref{fig:profile}. The narrow central 
component is the cause of the excellent timing precision for this pulsar.
This profile does not show significant changes with radio
frequency beyond what one should expect from the change in
time resolution of the system $dt$ with frequency. For this
reason, the best profile is then cross-correlated with {\em all}
the 4-minute/25 MHz  pulse profiles in the Fourier domain \citep{tay92};
the phase offset that yields the 
best match is used to derive the time of arrival (TOA) of a
particular pulse (normally that closest to the start of each
sub-integration) measured at the local (topocentric) time frame.
In total we obtain 17376 TOAs from Arecibo data. The previous number
of Parkes measurements is 100.

We carry out subsequent TOA analysis using the {\sc tempo2} software
package \citep{hem06,ehm06}.
This uses the clock corrections intrinsic to the observatory, the
Earth rotation data and the observatory coordinates to convert the
topocentric TOAs to Terrestrial Time (TT),
as maintained by the Bureau International des Poids et Mesures.
From these {\sc tempo2} derives arrival times at the Solar
System Barycentre using the position of the Observatory
that is calculated using the DE/LE~421 Solar System ephemeris \citep{fwb08}; 
for details see \cite{ehm06}. These, like our timing parameters, are expressed 
in Barycentric Coordinate Time (TCB).

With the correct rotation count, {\sc tempo2} can vary the timing
parameters in order to minimize the difference between the observed
and predicted barycentric TOAs (the timing residuals). The residuals
are weighted according to the estimated uncertainty of each TOA. The
resulting set of timing parameters constitutes a preliminary
phase-coherent timing solution (an ``ephemeris''). If the TOA
uncertainties are under-estimated, or if there are systematics
present in the data, the reduced $\chi^2$ of the fit will be
larger than 1. For this preliminary ephemeris, the reduced $\chi^2$
is 2.05, which, as we show below, is mostly caused by variations
in the intervening electron column density.

The main advantage of having WAPP search data stored on disk
and doing the analysis off-line is that it allows for iterative
improvement of the pulsar ephemeris; this process is
particularly important at the earlier stages when the ephemeris is
not yet very precise. With each improved ephemeris, we
dedispersed and folded the data again, obtaining pulse profiles with
improved signal-to-noise ratio; which are then used to derive better
TOAs that further improve the ephemeris. This avoids orbital
phase-dependent smearing of the pulse profiles and the creation of
orbital phase-dependent timing artefacts, which can corrupt the
determination of orbital parameters, particularly
the orbital phase and orbital period variation \citep{nsk08}.

At the early stages of our timing programme, we checked the
timing accuracy of the WAPPs by making a few simultaneous observations
with the Arecibo Signal Processor (ASP). The ASP is a real-time
coherent dedispersion system implemented in software, which
can process 64 MHz of baseband signal \citep{dem07} and is known to provide very 
stable timing. No significant differences in the timing of the two back-ends was
found, this implies that, within measurement precision, the WAPP timing is 
accurate.

%==============================================================================%

\subsection{DM variations and the timing model}
\label{sec:timing_model}

The preliminary ephemeris described above assumes a constant DM. As we
can see in the right plot of Fig.~\ref{fig:residuals}, this ephemeris
describes the TOAs rather poorly, with large trends in the post-fit
residuals. These are especially noticeable once we calculate daily
residual averages.

We used this preliminary ephemeris to measure the pulsar's DM for each
day of observations. This is only possible because of
the wide band of the L-wide receiver at Arecibo --- from 1120 to 1730
MHz, of which we used the cleanest parts, 1120-1220 MHz and 1260-1560 MHz.
The results of these measurements are presented in
the left plot of Fig.~\ref{fig:residuals}. The DM varies significantly
with time, and its variation correlates with the daily residual
averages; this implies that the latter must, to a large extent, be
caused by the former. They have amplitudes of the order of
0.002~cm$^{-3}$ pc, which cause extra dispersive delays of the order of
4~$\mu$s, and a dominant timescale of the order of a few years.
This is similar to the timescales associated with
GWs from super massive black hole
binaries. This highlights the importance of
accurate and dense multi-frequency TOA measurements
and precise DM corrections when attempting to detect
such waves with pulsar timing \citep{vbc+09}.

These DM variations cannot be described by a simple linear variation with
time. For this reason, we included eight DM derivatives in our definitive
timing model, which is presented in Table~\ref{tab:parameters}; the
uncertainties were derived by {\sc tempo2} except where stated
otherwise. This DM model describes the daily DM measurements
very well (see left plot of Fig.~\ref{fig:residuals}); 
but is only valid within the time interval for
which we have significant multi-frequency data, i.e., since the start
of the Arecibo observations in 2003.
The predicted values start diverging fast outside this range,
but this does not affect the results discussed in this paper.
This method has the benefit that any possible correlations
of the coefficients of the DM model with the timing parameters are
taken into account, leading to more conservative (and in our opinion
more realistic) estimates of the parameter uncertainties.

This ephemeris describes the TOAs much better than the preliminary
ephemeris, as displayed in the left plot of
Fig.~\ref{fig:residuals_corrected}. However, there are still
systematic variations in the daily residual averages, these excursions
reach a maximum of about 0.5 $\mu$s and have timescales
of a few months to a
year. In order to estimate the timing parameters with more realistic
uncertainties, we added an uncertainty of 0.5 $\mu$s in quadrature to
the estimated uncertainty of every TOA.

This rescaling of our errors results in a reduced $\chi^2$ of 1
for short timescales and 1.02 for the whole data set, suggesting that
the uncertainties on the derived parameters are reliable.
However, given the systematic nature of these excursions, small
systematics on parameters with comparable timescales (in particular
the parallax) may remain. On the other hand, there should be no
correlation with orbital phase, hence
orbital parameters should not be affected at all.

These excursions are likely to be caused by variations of the DM
at timescales of a few months
that cannot be taken into account by the 8-polynomial model for the DM
variations. The alternative explanations are not as plausible: An
instability in the rotation of the pulsar would produce excursions with
larger timescales than those observed. The second derivative of the spin
frequency [$\ddot{\nu} = (-0.6 \pm 2.3) \times 10^{-28}\,\rm
Hz\,s^{-2}$] is consistent with zero, which suggests good long-term
stability; the same is true for the third frequency derivative.
Furthermore, the agreement between nearby daily residual averages
suggests that the timing system used is stable.

To further verify the integrity of this timing solution, we compared its
predictions with TOAs taken with the Westerbork Synthesis Radio
Telescope (WSRT) in the Netherlands, which uses the Puma II coherent
dedispersion back-end \citep{kss08}. The 345 MHz data are especially useful 
because, in case of an inaccuracy in our DM model they would show significant
trends. It is remarkable that, despite the fact that the model
represents mostly an interpolation during the 2009--2010 gap, no trends
are discernible in either dataset (see right plot of
Fig.~\ref{fig:residuals_corrected}).

%==============================================================================%

\subsection{Orbital model}
\label{sec:orbital_model}

\begin{figure*}
  \includegraphics[width=17.5cm]{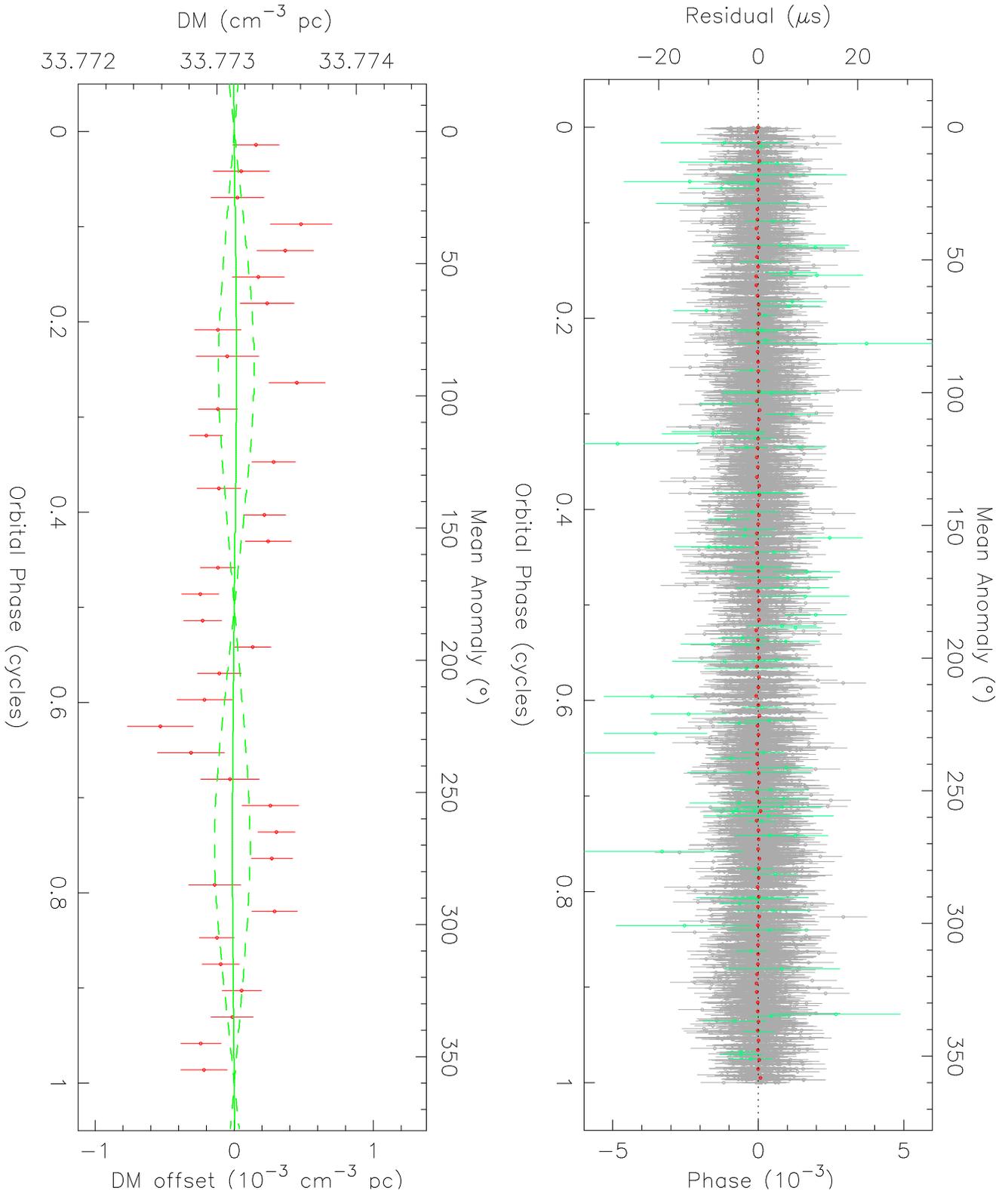}
  \caption{
  {\em Left}: Measured DM versus orbital phase. To the observed DMs
    we fit a DM variation model (green
    lines, see Section~\ref{sec:mass_loss}) using DM and $\Delta \rm DM$
    as the free parameters, and display  the resulting fits for the
    the nominal and $\pm 2$-$\sigma$ values of $\Delta \rm DM$.
    No significant variation as a function of orbital phase is
    detectable; indicating the lack of spurious artefacts in the data
    reduction and no outgassing from the companion.
   {\em Right:} Post-fit residuals versus orbital phase, which for
    this very low-eccentricity system is measured from ascending
    node (i.e., the mean anomaly is equal to the orbital longitude).
    The residuals of the Parkes (green) and Arecibo (gray)
    TOAs were obtained with the timing model presented in 
    Table~\ref{tab:parameters}.
    In red we indicate bin averages where the bin width is 0.01 $P_b$.
    No significant trends can be identified
    in the residuals or their binned averages; indicating that the
    orbital model can describe the orbital modulation of the TOAs correctly.
    No dispersive delays or unnacounted Shapiro delay signatures
    are detectable near orbital phase 0.25, nor
    artefacts due to incorrect dedispersion and folding of the data.    
  \label{fig:orbit}}
\end{figure*}

%-------------------------------------------------------------------------------

\begin{table*}
\begin{center}
\begin{tabular}{l l}
  \hline
  \multicolumn{2}{l}{{\bf Timing parameters}} \\
  Reference Time (MJD)              \dotfill & 54600.0001776275 \\
  Right Ascension, $\alpha$ (J2000) \dotfill & 17$^{\rm h}\;$38$^{\rm m}\;$53$\fs$9658386(7) \\
  Declination, $\delta$ (J2000)     \dotfill & 03$\degr\;$33$\arcmin\;$10$\farcs$86667(3)  \\
  Proper Motion in $\alpha$, $\mu_{\alpha}$ (mas yr$^{-1}$) \dotfill & $+$7.037(5)  \\
  Proper Motion in $\delta$, $\mu_{\delta}$ (mas yr$^{-1}$) \dotfill & $+$5.073(12) \\
  Parallax, $\pi_x$ (mas) (a)\dotfill & 0.68(5)\\
  Spin Frequency, $\nu$ (Hz)                  \dotfill &  170.93736991146392(3) \\
  First Derivative of $\nu$, $\dot{\nu}$ (fHz s$^{-1}$) \dotfill & $-$0.704774(4)  \\
  Orbital Period $P_b$ (days)                 \dotfill & 0.3547907398724(13) \\
  Projected Semi-Major Axis, $x$ (lt-s)       \dotfill & 0.343429130(17) \\
  Time of Ascending Node, $T_{\rm asc}$ (MJD) \dotfill & 54600.20040012(5)  \\
  $\eta \equiv e \sin \omega $ \dotfill & $(-1.4  \pm 1.1) \times 10^{-7}$  \\
  $\kappa \equiv e \cos \omega $ \dotfill & $(3.1 \pm 1.1) \times 10^{-7}$  \\
  First Derivative of $P_b$, $\dot{P}_b$ ($\rm fs\, s^{-1}$) \dotfill & $-$17.0(3.1)\\
  ``range" parameter of Shapiro delay, $r$ ($\mu$s) (b) \dotfill & 0.8915 \\
  ``shape" parameter of Sapiro delay, $s \equiv \sin i$ (b) \dotfill & 0.53877\\
  Dispersion Measure, DM (cm$^{-3}$ pc)       \dotfill & 33.77312(4)  \\
  First Derivative of DM, DM1 (cm$^{-3}$ pc yr$^{-1}$) \dotfill & +0.00108(3)  \\
  ... DM2 (cm$^{-3}$ pc yr$^{-2}$) \dotfill & $-$0.00041(3)  \\
  ... DM3 (cm$^{-3}$ pc yr$^{-3}$) \dotfill  & $-$0.000279(9)  \\
  ... DM4 (cm$^{-3}$ pc yr$^{-4}$) \dotfill & +0.000059(6)  \\
  ... DM5 (cm$^{-3}$ pc yr$^{-5}$) \dotfill & +0.0000230(19)  \\
  ... DM6 (cm$^{-3}$ pc yr$^{-6}$) \dotfill & $-$0.0000019(3)  \\
  ... DM7 (cm$^{-3}$ pc yr$^{-7}$) \dotfill & $-$0.00000090(11) \\
  ... DM8 (cm$^{-3}$ pc yr$^{-8}$) \dotfill & $-$0.000000060(10) \\[2mm]
  \multicolumn{2}{l}{{\it Test parameters}} \\
  First Derivative of $x$, $\dot{x}$ ($\rm fs\,s^{-1}$) \dotfill & 0.7(5)\\
  Second Derivative of $\nu$, $\ddot{\nu}$ ($10^{-28}$ Hz s$^{-2}$) \dotfill & $-$0.6(2.3) \\[2mm]
 
  \multicolumn{2}{l}{{\bf Optical Parameters}} \\
 
  Companion Mass, $M_c$ ($\rm M_{\odot}$) \dotfill & $0.181^{+0.008}_{-0.007}$  \\
  Spectroscopic Companion Radius, $R_c$  ($\rm R_{\odot}$) \dotfill & $0.037^{+0.004}_{-0.003}$  \\
  Semi-amplitude of orbital velocity of companion, $K_{c}$ ($\rm km \, s^{-1}$) \dotfill & 171(5) \\[2mm]

  \multicolumn{2}{l}{{\bf Derived Parameters}} \\

  Galactic Longitude, $l$ \dotfill & 27\fdg7213\\ 
  Galactic Latitude, $b$ \dotfill & 17\fdg7422\\
  Distance, $d$ (kpc) \dotfill & 1.47(10) \\
  Total Proper Motion, $\mu$ (mas yr$^{-1}$)  \dotfill & 8.675(8)\\
  Position angle of proper motion, $\Theta_{\mu}$ (J2000) \dotfill & 53\fdg72(7)\\
  Position angle of proper motion, $\Theta_{\mu}$ (Galactic) \dotfill & 116\fdg12(7)\\
  Spin Period, $P$ (s) \dotfill & 0.005850095859775683(5) \\
  First Derivative of Spin Period, $\dot{P}$ ($10^{-20} \rm s\, s^{-1}$) \dotfill & 2.411991(14) \\
  Intrinsic $\dot{P}$, $\dot{P}_{\rm Int}$ ($10^{-20} \rm s\, s^{-1}$) (a) \dotfill & 2.243(13)\\
  Characteristic Age, $\tau_c$ (Gyr)  \dotfill & 4.1 \\
  Transverse magnetic field at the poles, $B_0$ ($10^9$G) \dotfill & 0.37\\
  Rate or rotational energy loss, $\dot{E}$ ($10^{33}$ erg s$^{-1}$) \dotfill & 4.4 \\
  Mass Function, $f$ ($\rm M_{\odot}$)   \dotfill &  0.0003455012(11) \\
  Mass ratio, $q \equiv M_{p}/M_{c}$ \dotfill & 8.1(2) \\
  Orbital inclination, $i$ ($^\circ$) \dotfill & 32.6(1.0)\\
  Pulsar Mass, $M_p$ ($\rm M_{\odot}$) \dotfill & $1.46^{+0.06}_{-0.05}$\\
  Total Mass of Binary, $M_t$ ($\rm M_{\odot}$) \dotfill & 1.65$^{+0.07}_{-0.06}$\\    
  Eccentricity, $e$ \dotfill &  $(3.4 \pm 1.1) \times 10^{-7}$ \\
  Apparent $\dot{P}_b$ due to Shklovskii effect, $\dot{P}_b^{\rm Shk}$
  ($\rm fs \, s^{-1}$) (a)\dotfill & $8.2^{+0.6}_{-0.5}$\\
  Apparent $\dot{P}_b$ due to Galactic acceleration, $\dot{P}_b^{\rm Gal}$
  ($\rm fs \, s^{-1}$) (a) \dotfill & $0.58^{+0.16}_{-0.14}$\\
  Intrinsic $\dot{P}_b$, $\dot{P}_b^{\rm Int}$
  ($\rm fs \, s^{-1}$) (a) \dotfill & $-$25.9(3.2)\\
  Predicted $\dot{P}_b$, $\dot{P}_b^{\rm GR}$
  ($\rm fs \, s^{-1}$) \dotfill & $-$27.7$^{+1.5}_{-1.9}$\\
  ``Excess'' orbital decay, $\dot{P}_b^{\rm xs} =
  \dot{P}_b^{\rm Int}-\dot{P}_b^{\rm GR}$ ($\rm fs \, s^{-1}$) (a) \dotfill & $ +2.0^{+3.7}_{-3.6}$ \\
  Time until coalescence, $\tau_m$ (Gyr) \dotfill & $\sim$ 13.2 \\
  \hline
\end{tabular}
\caption{\label{tab:parameters}
  Parameters for the \psr\ system. In parentheses we present the 1-$\sigma$
  uncertainties in the last digit quoted, as estimated by {\sc tempo2}. If the 
  value and uncertainty are signaled with an (a) then they were derived from a 
  Monte-Carlo procedure (Section~\ref{sec:MonteCarlo}). (b) The Shapiro delay 
  parameters $r$ and $s$ were not fitted in the derivation of the timing model;
  the values used were derived from a combination of other timing and optical 
  parameters (Section~\ref{sec:orbital_model}). All timing parameters are 
  derived using {\sc tempo2} and are displayed as measured at the Solar System 
  Barycenter, in barycentric coordinate time (TCB). The ``test parameters" were 
  not fitted when deriving the main timing model, but their values were derived 
  fitting for all the other parameters in the model.}
\end{center}
\end{table*}

%-------------------------------------------------------------------------------

The orbit of \psr\ has a very low eccentricity, so we use the ``ELL1''
orbital model \citep{lcw+01} to model it.\footnote{The ELL1 timing model as 
implemented in the TEMPO software package 
(http://sourceforge.net/projects/tempo/) is a modification of the DD timing 
model \citep{dd85,dd86} adapted to low-eccentricity binary pulsars.
In terms of  
post-Keplerian observables, it contains all those which are numerically
relevant for systems with $e \ll 1$. The ``Einstein delay" term is
not relevant for such systems and is therefore not taken into account.}
This yields Keplerian and
post-Keplerian parameters that are very weakly correlated with each
other. Note that, in order to estimate the ``real'' eccentricity of the
binary we assumed that the Shapiro delay is as predicted by general
relativity for the values of $M_c$ and $\sin i$ derived in Paper I. This
assumption is safe because GR is known to provide a
sufficiently accurate description of the distortion of space-time around weakly
self-gravitating objects \citep{bit03}.

According to \cite{fw10}, the orthometric
amplitude of the Shapiro delay (which quantifies the time amplitude of
the {\em measurable} part of the Shapiro delay) is, for this system,
given by $h_3\,=\,22\,$ns. Fitting for this quantity we obtain $h_3 =
9\,\pm\, 13\,$ns. This is 1-$\sigma$ consistent with the prediction but
the low relative precision of this measurement implies that we cannot
determine $M_c$ and $\sin i$ independently from the existing timing data.
A precise measurement of the component masses of this system from
Shapiro delay would require an improvement in timing precision that
is much beyond our current capabilities.

In the right plot of Fig.~\ref{fig:orbit}, we display the residuals as a
function of the orbital phase. No trends are noticeable, either in the
residuals or their averages, this implies that the orbital model is
not obviously flawed. This also suggests that the timing system is
inherently stable. Furthermore, we see no DM variations as a function
of the orbital phase (left plot of Fig.~\ref{fig:orbit}); i.e.,
there are no obvious spurious DM artefacts caused by incorrect folding
nor detectable dispersive delays in the data.

%%%%%%%%%%%%%%%%%%%%%%%%%%%%%%%%%%%%%%%%%%%%%%%%%%%%%%%%%%%%%%%%%%%%%%%%%%%%%%%%

\section{Results}
\label{sec:results}

In what follows, we discuss some of the results of the timing program.
In Section~\ref{sec:parallax} we briefly discuss the measurement
of the parallax, which requires special care given the systematics
highlighted in Section~\ref{sec:timing_model}.
Then in Section~\ref{sec:intrinsic_decay} we focus on the main result of
the timing: the detection of the orbital decay of the system, $\dot{P}_b$.
We compare it with the GR prediction in Section~\ref{sec:excess_decay}.

%==============================================================================%

\subsection{Parallax}
\label{sec:parallax}

As mentioned in Section~\ref{sec:timing_model}, the quoted uncertainty
for the parallax is likely to be too small given the systematic
effects caused by uncorrected short-term DM variations. It is
therefore important to gain a sense of whether this parallax estimate
is accurate or if there are inconsistencies with other distance
estimators.

The distance we obtain from this parallax ($d\,=\,1.47\pm
0.10\,$kpc) is consistent with the 1.4 kpc predicted by the NE2001 electron
model of the Galaxy \citep{cl01} for the pulsar's Galactic coordinates
and DM. However, the distance estimates based on this model have been
shown, in some cases, to significantly under- or overestimate real distances,
therefore this agreement with the DM prediction cannot be used as solid
evidence that our parallax is accurate.

This distance, when combined with the known temperature and photometric
properties of the white dwarf, produces an estimate for its
radius that is very similar to that derived from its the spectrum
(see Paper I). This suggests that our value for the distance is likely to
be accurate given the present timing uncertainties. This is important
because the distance (and the proper motion, also presented
in Table~\ref{tab:parameters}) are necessary for a correct estimate of the
intrinsic orbital decay of the system $\dot{P}_b^{\rm Int}$,
as discussed below.

Following the analysis by \cite{vlm10}, we find that there are no significant 
biases affecting this parallax measurement.

%==============================================================================%

\subsection{Intrinsic orbital decay}
\label{sec:intrinsic_decay}

The intrinsic orbital decay of the system can be obtained from the
observed orbital period variation ($\dot{P}_b$) by subtracting the
{\em kinematic effects} \citep{shk70,dt91}:
\begin{equation}
\label{eq:intrinsic_pbdot}
\dot{P}_b^{\rm Int} = \dot{P}_b -
\dot{P}_b^{\rm Acc} - \dot{P}_b^{\rm Shk}.
\end{equation}
The same equation applies to any quantity with the
dimension of time, like the spin parameters ($P$, $\dot{P}$
and $\dot{P}^{\rm Int}$).

The first term,
$\dot{P}_b^{\rm Acc} = a_T P_b / c$ is caused by the difference
of accelerations of the \psr\ system and the Solar System
projected along the line of sight to the pulsar, $a_T$.
This term is dominated by the difference of accelerations of the two
systems caused by the average Galactic field, $a_G$:
For the pulsar's Galactic coordinates of $l =  27\fdg7213$
and $b = 17\fdg7422$ and distance, we
obtain [using eq.~(5) in \cite{nt95}, in
combination with eq.~(17) in Lazaridis~et~al.~2009]
\begin{equation}
  \frac{a_G}{c} P_b = 0.58^{+0.16}_{-0.14} \rm \, fs \,s^{-1}.
\end{equation}
The $a_T$ term also contains a contribution from
nearby masses, $a_S$. Damour \& Taylor (1991) present a statistical
estimate of the magnitude of this effect for PSR B1913+16, which
is dominated by large molecular clouds. For \psr, the same $a_S$
would yield an orbital period derivative of
\begin{equation}
\left| \frac{a_S}{c} \right| P_b \lesssim 0.2 \rm \, fs \,s^{-1},
\end{equation}
which is negligible. The $a_S$ of \psr\ is likely to be
even smaller: although the pulsar is at a similar Galactocentric
distance as PSR~B1913+16,
it is at a Galactic height of $\sim\,0.45\,$kpc, which is larger
than that of PSR~B1913+16: $\sim \, 0.3\, $kpc.

The second kinematic effect in eq.~(\ref{eq:intrinsic_pbdot}), commonly known as 
the ``Shklovskii'' effect, caused by the centrifugal acceleration, is given by \citep{shk70}:
\begin{equation}\label{eq:shk}
  \dot{P}_b^{\rm Shk} 
    = \left( \mu_{\alpha}^{2} + \mu_{\delta}^{2} \right) \frac{d}{c} P_b
    = 8.3^{+0.6}_{-0.5} \rm \, fs \, s^{-1},
\end{equation}
where $\mu_{\alpha}$ and 
$\mu_{\delta}$ are the proper motion in right
ascension and declination (see Table~\ref{tab:parameters}).
For the intrinsic $\dot{P}_b$, we thus obtain
\begin{equation}\label{eq:PbdotInt}
  \dot{P}_b^{\rm Int}  = -25.9 \pm 3.2\rm \, fs \, s^{-1} .
\end{equation}
%%

%==============================================================================%

\subsection{Excess orbital decay}
\label{sec:excess_decay}

%----------

\begin{figure*}
  \includegraphics[width=16.7cm]{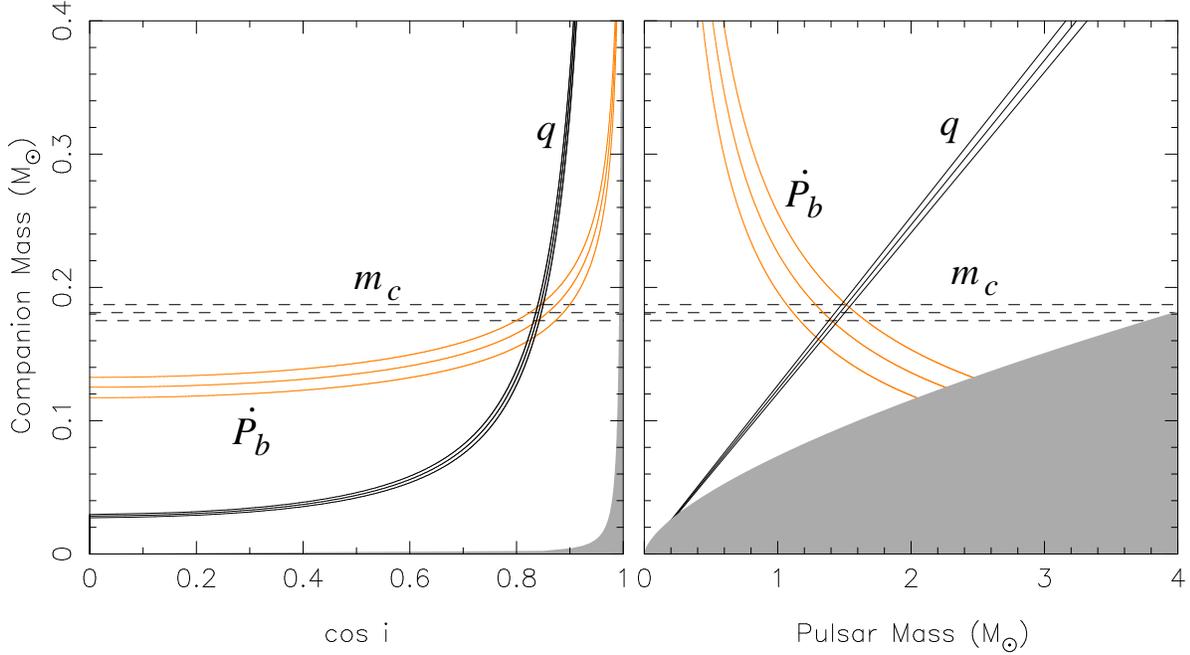}
  \caption{Constraints on system masses and orbital inclination from
    radio and optical measurements of \psr\ and its WD companion.
    The mass ratio $q$ and the companion mass $m_c$ are theory-independent
    (indicated in black), but the constraints from the measured
    intrinsic orbital decay
    ($\dot{P}_b^{\rm Int}$, in orange) are calculated {\em assuming} that
    GR is the correct theory of gravity. All curves intersect, meaning
    that GR passes this important test. {\em Left}: $\cos i$--$m_c$
    plane. The gray region is excluded by the condition $m_p >
    0$. {\em Right}: $m_p$--$m_c$ plane. The gray region is excluded
    by the condition $\sin i \leq 1$.  Each triplet of curves corresponds
    to the most likely value and standard deviations of the respective
    parameters.
  \label{fig:mass_mass}}
\end{figure*}

%----------

From the values for $q$ and $m_c$ in Paper I we can
estimate the orbital decay caused by the emission of
quadrupolar GWs for a low-eccentricity system, as
predicted by GR:
\begin{eqnarray}\label{eq:gr}
  \dot{P}_b^{\rm GR} & \simeq & -\frac{192\,\pi}{5}
    \left(n_b {\rm T_{\odot}}\,m_{c} \right)^{5/3}
    \frac{q}{(q+1)^{1/3}} \nonumber\\
   & = & -27.7^{+1.5}_{-1.9} \rm \, fs \, s^{-1}, 
\end{eqnarray}
where ${\rm T_{\odot}} \equiv G {\rm M_{\odot}} c^{-3} = 4.925 490 947\,\mu$s
\citep{lk05}. Subtracting this from $\dot{P}_b^{\rm Int}$ 
(eq.~(\ref{eq:PbdotInt})) we obtain the ``excess'' orbital decay relative to the 
prediction of GR,
\begin{equation}\label{eq:PbdotEx}
  \dot{P}_b^{\rm xs} = 2.0^{+3.7}_{-3.6} \rm \, fs \, s^{-1}.
\end{equation}
This is consistent with zero. As discussed in Section~\ref{sec:generic},
this implies that
GR passes the test posed by the orbital decay
of \psr. We illustrate this match in Fig.~\ref{fig:mass_mass}, where
we see that the mass/inclination constraints, derived from
$\dot{P}_b^{\rm Int}$ using eq.~(\ref{eq:gr})
(i.e., assuming that GR is
the correct theory of gravity), are consistent with the
theory-independent\footnote{By {\em theory-independent}, in the context of this paper, we denote quantities which are either based on weak-field gravity, which is known to be described extremely well by GR \citep{will06}, or quantities which are free of any explicit strong field deviations of gravity from GR, at least within the wide class of Lorentz-invariant gravity theories. The mass ratio in a binary pulsar system is an example of the latter \citep{dam07}.}
constraints derived from the optical observations.

%==============================================================================%

\subsection{Rigorous uncertainty estimates for orbital decay}
\label{sec:MonteCarlo}

To make reliable estimates of the uncertainties of these derived
quantities (${\dot P}_b^{\rm Gal}$, $\dot{P}_b^{\rm Shk}$,
$\dot{P}_b^{\rm Int}$ and $\dot{P}_b^{\rm xs}$), we implemented a
new Monte Carlo routine in {\sc tempo2}. From our list of 17476 TOAs,
we created 220\,000 similar data sets of fake TOAs that have random 
distributions
consistent with the original TOAs and their uncertainties. For each
fake TOA data-set, we run {\sc tempo2} and record the resulting
best-fit parameters to disk. We then use the parallax of
each simulation to estimate $a_G$ and calculate the
derived quantities for that simulation
using the equations above. Finally, in order to
estimate $\dot{P}_b^{\rm xs}$, we use a random ($m_c, q$) pair
from the Monte Carlo simulation in Paper I, calculate
the corresponding  $\dot{P}_b^{\rm GR}$ and subtract this
from $\dot{P}_b^{\rm Int}$.
This procedure is warranted by the intrinsic lack of correlation between
the optical and radio measurements. The computer then calculates
averages, standard deviations, medians and $\pm 1$-$\sigma$
percentiles (presented in Table~\ref{tab:parameters}) for the
resulting distributions of derived quantities. The averages and standard
deviations are close to the estimated medians and 1-$\sigma$
percentiles, implying that the resulting distributions are generally
close to Gaussian.

With this method, we are able to take into account any underlying
correlations between the observables, thus estimating more reliable
values and uncertainties for the derived quantities
that depend only on the measured TOAs and their uncertainties.

%%%%%%%%%%%%%%%%%%%%%%%%%%%%%%%%%%%%%%%%%%%%%%%%%%%%%%%%%%%%%%%%%%%%%%%%%%%%%%%%

\section{Generic tests of gravity theories}
\label{sec:generic}

In order to understand the significance of the small value of $\dot{P}_b^{\rm
xs}$ in eq.~(\ref{eq:PbdotEx}) --- the main experimental result of this paper 
--- we now discuss what physical effects could in principle be contributing to 
it. According to \cite{dt91}
\begin{equation} \label{eq:xs_contributions} 
\dot{P}_b^{\rm xs} = \dot{P}_b^{\dot{M}} + \dot{P}_b^{\rm T} +
                     \dot{P}_b^{\rm D} + \dot{P}_b^{\dot{G}} ,
\end{equation} 
where $\dot{P}_b^{\dot{M}}$ is due to mass loss from the binary,
$\dot{P}_b^{\rm T}$ is a contribution from tidal effects,
$\dot{P}_b^{\rm D}$ is the orbital decay caused mainly
by the emission of dipolar GWs (and any extra
multipole modifying the general relativistic prediction)
and $\dot{P}_b^{\dot{G}}$ is a contribution from possible (yet undetected)
variations of Newton's gravitational constant (as measured by a Cavendish
experiment).
The first two terms are the ``classical" terms, the
last two would only be non-zero for theories of gravity other than general
relativity.

%==============================================================================%

\subsection{Classical terms}

\subsubsection{Mass loss}

In Appendix~\ref{sec:mass_loss}, we derive an upper limit for the
mass loss from the companion as a function of the total mass
of the system. For the pulsar, the mass loss is dominated
by the loss of rotational energy \citep{dt91}:
\begin{equation}
  \frac{\dot{M}_p}{M_t} 
    = \frac{\dot{E}}{M_t c^2} 
    = 1.5 \times 10^{-21} \, {\rm s}^{-1},
\end{equation}
which is of the same order as the upper limit for $\frac{\dot{M}_c}{M_t}$.

The contribution to the orbital variation due to the total
mass loss $\dot{M} = \dot{M}_c + \dot{M}_p$
is given by \citep{dt91}:
\begin{equation}
\label{eq:massloss}
\dot{P}_{b}^{\dot{M}} = 2 \frac{\dot{M}}{M_t} P_b
                    < 0.2 \rm\, fs \, s^{-1}, 
\end{equation}
which is about 20 times smaller than the current uncertainty in the
measurement of $\dot{P}_b^{\rm xs}$.

\subsubsection{Tidal orbital decay}

We now calculate the orbital decay caused by tides. From
eqs. (3.15) and (3.19) in \cite{sb76}, we derive the
following  expression for $\dot{P}_b^{\rm T}$:
\begin{equation}
\dot{P}_b^{\rm T} = \frac{k \Omega_c}{3 \pi q (q +1)} 
\left( \frac{R_c P_b \sin i}{x c} \right)^2 \frac{1}{\tau_s}.
\end{equation}
Unlike the expressions in eq. (3.19), this equation
is exact because it relates the synchronisation
timescale $\tau_s$ (which describes the change in the
companion angular velocity
$\Omega_c$, $\tau_s = - \Omega_c / \dot{\Omega}_c$) to
the timescale associated with the change in the orbital
period ($\tau_p = P_b / \dot{P}_b^{\rm T}$)
assuming only conservation of the angular momentum. In
this expression $k \equiv I_c / (M_c R_c^2)$, where $I_c$ is the
WD moment of inertia. White dwarfs (particularly those
with a mass much below the Chandrasekhar limit)
are sustained by the degeneracy pressure of non-relativistic electrons and
can be well approximated by a polytropic sphere with $n = 1.5$.
For such stars, we have $k = 0.2$ \citep{motz52}.

The only unknown parameters in this expression are $\Omega_c$ and $\tau_s$.
If $\tau_s$ is much smaller than the characteristic age of the pulsar 
$\tau_c = 4.1\,$Gyr, then the WD rotation is already synchronised with the orbit 
($\Omega_c = n_b$) and there are no tidal effects at all. If, on the other hand, 
$\tau_s > \tau_c$, then $\Omega_c$ can be much larger, but it must still be 
smaller than the break-up angular velocity $\Omega_c < \sqrt{G M_c / R_c^3} = 
0.038\, \rm rad\, s^{-1}$. These conditions for $\Omega_c$ and $\tau_s$ yield 
$\dot{P}_b^{\rm T} < 1.4 \,\rm fs \, s^{-1}$. Thus, even if the WD were rotating 
near break-up velocity, $\dot{P}_b^{\rm T}$ would still be smaller than the 
uncertainty in the measurement of $\dot{P}_b^{\rm xs}$. We note, however, that 
the progenitor of the WD was very likely synchronised with the orbit. This 
implies that, when the WD formed, its rotational frequency was within one order 
of magnitude of the orbital frequency, i.e., $\Omega_c \lesssim 2 \times 10^{-3} \, 
\rm rad \, s^{-1}$ (for the reasoning, see, e.g., Appendix B2.2 of 
\cite{bkkv06}).

%==============================================================================%

\subsection{Test of GR and generic tests of alternative gravity theories}
\label{sec:generic_tests}

%----------

\begin{figure}
  \includegraphics[width=8.5cm]{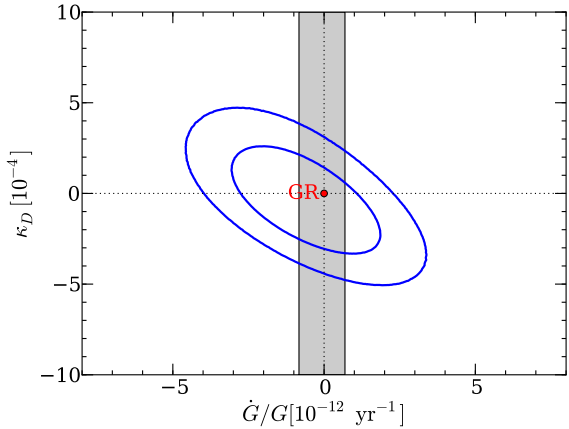}
  \caption{Limits on $\dot{G}/G$ and $\kappa_D$ derived from the
    measurements of $\dot{P}_b^{\rm xs}$ of \psr\ and
    PSR~J0437$-$4715. The inner blue contour level includes 68.3\%
    and the outer contour level 95.4\% of all probability.
    At the origin of coordinates, general
    relativity is well within the inner contour and close to the peak
    of probability density. The gray band includes regions
    consistent with the measured value and 1-$\sigma$ 
    uncertainty of $\dot{G}/G$ from Lunar Laser Ranging (LLR). Generally 
    only the upper half of the diagram has physical meaning, as the
    radiation of dipolar GWs must necessarily make the
    system lose orbital energy.
  \label{fig:generic_tests}}
\end{figure}

%----------

The smallness of the classical terms implies that the measurement of 
$\dot{P}_b^{\rm xs}$ (eq.~(\ref{eq:PbdotEx})) is a direct test of GR. Unlike 
many alternative theories of gravity, GR predicts $\dot{P}_b^{\dot{G}} = 0$, 
$\dot{P}_b^{\rm D} = 0$ and therefore $\dot{P}_b^{\rm xs} = 0$. As discussed in 
Section~\ref{sec:excess_decay}, this is consistent with observations, which 
means that GR passes the test posed by the measurement of $\dot{P}_b^{\rm xs}$ 
for \psr. In this respect \psr\ constitutes a verification of GR's 
quadrupole formula with a precision of about 15\% at the 1-$\sigma$ level. 
In view of more stringent tests with other binary pulsars \citep{ksm+06,wnt10}
this result by itself does not seem particularly interesting. However, the large 
difference in the compactness of the two components of this binary system makes
\psr\ a remarkable laboratory for alternative gravity theories, in particular 
those which predict the emission of dipolar gravitational radiation. In 
Sections \ref{sec:STe} and \ref{sec:TeVeS}, we will confront our 
observations with two specific classes of gravity theories. In the present 
section, we follow a more generic approach, valid for gravity theories where 
nonperturbative strong-field effects are absent and higher-order contributions 
in powers of the gravitational binding energies of the bodies can be neglected, 
at least to a point where one does not care about multiplicative factors 
$\la 2$. As an example, the well known Jordan-Fierz-Brans-Dicke scalar-tensor 
theory falls into this group. 

Under the assumptions above, we can write for the change in the orbital period 
caused by dipolar gravitational radiation damping in a 
low-eccentricity binary pulsar system
\begin{equation}\label{eq:pbdotD}
\dot{P}_b^{\rm D} \simeq 
  - 2 \pi \, n_b{\rm T_{\odot}}m_c \, \frac{q}{q+1}\, \kappa_D {\cal S}^2
  + {\cal O}\left(s_{p,c}^3\right) \,,
\end{equation}
where ${\cal S} = s_p - s_c$ is the body-dependent term which is
given by the difference in the ``sensitivities'' of the 
pulsar, $s_p$, and the companion, $s_c$ [see \cite{wil93} for
the definition of $s_{p,c}$]. The quantity $\kappa_D$ is a body-independent 
constant that quantifies the dipolar self-gravity contribution, and takes 
different values for different theories of gravity.\footnote{In general 
scalar-tensor theories of gravity, we have $\kappa_D = 2 \eta^2
\left(1-\gamma^{\rm PPN}\right)^{-1}$, where $\eta \equiv 4 \beta^{\rm PPN} - 
\gamma^{\rm PPN} - 3$ is the Nordtvedt parameter, a combination 
of PPN parameters related to the violation of the strong equivalence principle.} 
For the purpose of this section, we have neglected higher-order corrections in 
powers of the sensitivities in the equation above (They actually vanish in the 
Jordan-Fierz-Brans-Dicke case). The full non-linearity will 
be taken into account in Sections \ref{sec:STe} and \ref{sec:TeVeS} 
(anticipating on the notation defined there, the terms $\propto s_{p,c}^3$ are 
negligible when the absolute value of the nonlinear matter-scalar coupling constant, $|\beta_0|$, is significantly smaller than 2). 

The value of $s_{p,c}$ depends on the theory of gravity, the exact form of the 
equation of state and the mass of the pulsar; for a $M_p = 1.4\,\rm M_{\odot}$ 
neutron star, it is generally of the order of 0.15, the value we use in 
our calculations. For an asymmetric system like \psr, the sensitivity of the 
companion WD has a negligible value: in the post-Newtonian limit, it is given by 
$\epsilon / M_c c^2 \sim 10^{-4}$, where $\epsilon$ is the gravitational binding 
energy of the WD \citep{will06}. Therefore, ${\cal S} = s_p - s_c \simeq s_p 
\neq 0$, which implies that if $\kappa_D \neq 0$, then there must be emission 
of dipolar GWs, and an associated orbital decay according to 
eq.~(\ref{eq:pbdotD}).
In a double neutron star system we would have $s_p \approx s_c$ and therefore 
${\cal S} \approx 0$, which means that we should observe $\dot{P}_b^{\rm D} 
\approx 0$ even if $\kappa_D \neq 0$. It is for this reason that, despite the 
low relative precision of the radiative test in \psr, it represents such a 
powerful constraint on alternative theories of gravity \citep[see, e.g.][]{ear75,bbv08}.
Apart from this, the use of optical data is very important because they provide 
estimates of $q$ and $m_c$ that are free of explicit strong field effects --- 
unlike in the case of the binary pulsar PSR~B1913+16 \citep[see][]{wnt10}, or 
for many of the parameters of the double pulsar \citep{ksm+06}. 

The last contribution to
$\dot{P}_b^{\rm xs}$ comes from a possible contribution to the orbital
change by a varying gravitational constant ($\dot{P}_b^{\dot{G}}$ in 
eq.~(\ref{eq:xs_contributions})). In the worst case, $\dot{P}_b^{\rm D}$ and $
\dot{P}_b^{\dot{G}}$ could both be large (in violation of GR)
but just happen to cancel each other 
in the \psr\ system because of different signs. To disentangle these effects 
there are two methods. First, one can use the best current limits from tests in 
the Solar System, notably Lunar Laser Ranging (LLR), which yields $\dot{G}/G = 
(-0.7 \pm 3.8) \times 10^{-13}$\,yr$^{-1}$ \citep{hmb10}, and obtain for \psr\ a (conservative) upper limit of
\begin{equation}
\dot{P}_b^{\dot{G}} 
   = -2 \frac{\dot{G}}{G} P_b 
   = (+0.14 \pm 0.74) \, \rm fs\, s^{-1} ,
\end{equation}
\citep{dgt88,dt91}. Therefore, $\dot{P}_b^{\rm D} = \dot{P}_b^{\rm xs} - 
\dot{P}_b^{\dot{G}} = 1.9_{-3.7}^{+3.8}\,{\rm fs\,s^{-1}}$, which yields, for a 
typical sensitivity $s_p = 0.15$
\begin{equation}
  \kappa_D = (-0.8 \pm 1.6) \times 10^{-4} \,,
\end{equation}
a limit that is a factor of eight more stringent than the limit from PSR~J1012+5307 \citep{lwj+09}.

The second method, developed in \cite{lwj+09}, 
combines two binary pulsar systems with different orbital periods. The method is 
based on the fact that a wide orbit is more sensitive to a change in the 
gravitational constant but less affected by the emission of dipolar GWs, in 
comparison to a more compact orbit. If we combine the $\dot{P}_b^{\rm xs}$ of
 \psr\ with that of a binary pulsar with a longer orbital period
we obtain a simultaneous test for $\kappa_D$ and $\dot{G}$.

When calculating $\dot{P}_b^{\dot{G}}$ for a combined limit on $\kappa_D$ and 
$\dot{G}$ based on two binary pulsars, we need to account for mass variations 
in compact stars as a result of a changing gravitational constant. Otherwise our 
limit on $\dot{G}$ will be too tight (Nordtvedt 1990). As a first 
approximation, that only accounts for the influence of the local value
of $G$, we can use eq.~(18) in Nordtvedt (1990):
\begin{equation}
\label{eq:pbdotGdot}
\dot{P}_b^{\dot{G}} = -2\,\frac{\dot{G}}{G} 
                   \left(1 - \frac{2q + 3}{2q + 2}\,s_p
                           - \frac{3q + 2}{2q + 2}\,s_c \right) P_{b} \,.
\end{equation}
As in eq.~(\ref{eq:pbdotD}), the contribution from the sensitivity of the 
white-dwarf companion, $s_c$, can be neglected.
For \psr, the correction factor due to the sensitivities (i.e., the parenthesis on
the right hand side of eq~(\ref{eq:pbdotGdot})) is about 0.85.

As in \cite{lwj+09}, we use the $\dot{P}_b^{\rm xs}$ of 
PSR~J0437$-$4715 \citep{dvtb08,vbs+08} to complement our $\dot{P}_b^{\rm xs}$ 
measurement (see eq.~(\ref{eq:PbdotEx})). 
PSR~J0437$-$4715 has a slightly higher mass than \psr,
and we will account for this in the sensitivity by having $s_p$ scale 
proportional to the mass, as suggested by eq.~(B.3) of \cite{de92}. The 
joint probability density function 
for $\dot{G}/G$ and $\kappa_D$ is displayed in Fig.~\ref{fig:generic_tests}. 
At the origin of coordinates, GR is well within the inner 68\% 
contour and close to the peak of probability density, i.e., it is consistent 
with the experimental results from these two binaries. Marginalizing this 
probability distribution function, we obtain
\begin{eqnarray}
  \dot{G}/G &=& (-0.6 \pm 1.6) \times 10^{-12}\,{\rm yr}^{-1} \nonumber\\
            &=& (-0.009 \pm 0.022) \, H_0, \\
  \kappa_D  &=& (-0.3 \pm 2.0) \times 10^{-4},
\end{eqnarray}
where $H_0$ is Hubble's constant \citep{rmc+09} and the uncertainties
are 1-$\sigma$. The $\dot{P}_b^{\rm xs}$ measurement of PSR~J0437$-$4715
is mostly responsible for the limit on $\dot{G}/G$, and it has
therefore not improved since \cite{lwj+09}. The
$\dot{P}_b^{\rm xs}$ measurement of \psr\ is mostly responsible
for the limit on $\kappa_D$, which has improved by a factor of $\sim 6$ 
since Lazaridis et al.~(2009). 
Although the limit on $\dot{G}/G$ derived from binary pulsar experiments is one 
order of magnitude less restrictive than that derived from LLR, it is of 
interest because it represents an independent test. 

The analysis presented in this section is restricted to gravity theories that 
do not develop nonperturbative strong-field effects in neutron stars. 
This assumption is well justified for \psr, since such effects do 
not seem to exist in other binary pulsars with similar masses, or even with a 
higher mass like in the case of PSR~J1012+5307 \citep{lwj+09}. 
Even when non-perturbative effects do develop, we will show
below that the higher-order corrections entering eq.~(\ref{eq:pbdotD})
do not change the conclusions qualitatively.

%%%%%%%%%%%%%%%%%%%%%%%%%%%%%%%%%%%%%%%%%%%%%%%%%%%%%%%%%%%%%%%%%%%%%%%%%%%%%%%%

\section{Constraints on scalar-tensor theories of gravity}
\label{sec:STe}

\begin{figure}
\includegraphics[width=8.5cm]{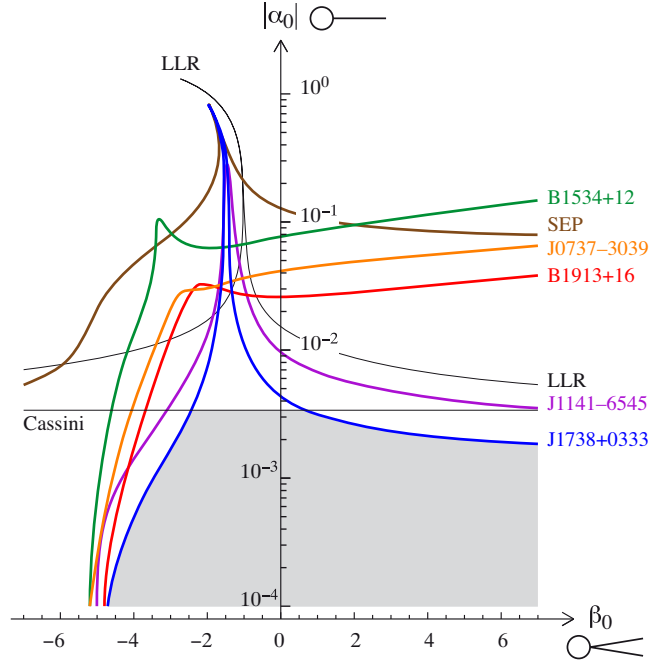}
\caption{\label{ScalarTensorPlane}
Solar-system and binary pulsar 1-$\sigma$ constraints
on the matter-scalar coupling 
constants $\alpha_0$ and $\beta_0$. Note that a logarithmic scale is used for 
the vertical axis $|\alpha_0|$, i.e., that GR ($\alpha_0 = 
\beta_0 = 0$) is sent at an infinite distance down this axis. LLR stands for 
lunar laser ranging, Cassini for the measurement of a Shapiro time-delay
variation in the Solar System, and SEP for tests of the strong 
equivalence principle using a set of neutron star-white dwarf low-eccentricity 
binaries (see text). The allowed region is shaded, and it includes general 
relativity. \psr\ is the most constraining binary pulsar, although the Cassini 
bound is still better for a finite range of quadratic coupling $\beta_0$.}
\end{figure}

The most natural alternatives to GR involve a
scalar field $\varphi$ contributing to the gravitational
interaction, in addition to the metric $g_{\mu\nu}$ describing
usual spin-2 gravitons. In these theories, matter is assumed to be
universally coupled to a physical metric $\tilde g_{\mu\nu}
\equiv A^2(\varphi) g_{\mu\nu}$, where $A(\varphi)$ is a
non-vanishing function defining the matter-scalar coupling.

It is convenient to expand this function around the background value 
$\varphi_0$ imposed by the cosmological evolution, as $\ln A(\varphi) = 
\ln A(\varphi_0)+ \alpha_0 (\varphi-\varphi_0) + \frac{1}{2} \beta_0
(\varphi - \varphi_0)^2 + \ldots$, where $\alpha_0$ defines the linear 
matter-scalar coupling constant, $\beta_0$ the quadratic coupling of matter to 
two scalar particles, and we will not consider higher-order vertices in the 
following (those corresponding to $\alpha_0$ and $\beta_0$ are diagrammatically
represented on the axes of Figs.~\ref{ScalarTensorPlane} and \ref{TeVeSPlane}, 
the circles meaning a matter source). GR corresponds to 
$\alpha_0 = \beta_0 = 0$, and Brans-Dicke theory \citep{jor59,fie56,bd61} to
$\alpha_0^2 =1/(2\omega_{\rm BD}+3)$ and $\beta_0 = 0$.

The predictions of such theories have been carefully studied in
the literature \citep{wil93,de92,de93,de96a,de96b,de98}. In
strong-field conditions, notably within and near a neutron star,
the coupling constants $\alpha_0$ and $\beta_0$ are modified by
self-gravity effects, and become body-dependent quantities,
$\alpha_A$ and $\beta_A$ ($A$ being a label for the body), which
can be computed by numerical integration of the field equations.
One needs to assume a specific equation of state (EOS) for
nuclear matter in such integrations, and we will use the moderate
one of \cite{de96b} in the following. The dependence on the stiffness of the EOS 
is illustrated in \cite{de98}, but this does not change the {\em relative} 
strength of the various binary-pulsar tests.

The body-dependent parameters $\alpha_A$ and $\beta_A$ enter all
observable predictions, and for instance, the effective
gravitational constant between two bodies $A$ and $B$ reads
\begin{equation}
  \tilde G_{AB} \equiv G_{\ast} A^2(\varphi_0)
  \cdot (1+\alpha_A \alpha_B),
\label{GAB}
\end{equation}
where $G_{\ast}$ denotes Newton's bare constant\footnote{Newton's constant $G$, 
as measured in the Cavendish experiment, is given by $G_{\ast} 
A^2(\varphi_0) \cdot (1 + \alpha_0^2)$.}. This induces a generic time dependence 
of the gravitational constant (cf.\ eq.~(\ref{eq:pbdotGdot})), as well as a 
violation of the strong equivalence principle (SEP): The acceleration of a body 
depends on its gravitational binding energy. Let us also quote the expressions
taken by the generalizations of Eddington's PPN parameters
$\gamma^\mathrm{PPN}$ and $\beta^\mathrm{PPN}$ \citep{edd23}, as
the first will differ in Section~\ref{sec:TeVeS} below:
\begin{eqnarray}
  \gamma_{AB} &\equiv&
    1 - 2 \frac{\alpha_A \alpha_B}{1+\alpha_A \alpha_B}, 
  \label{gammaAB}\\
  \beta^A_{BC} &\equiv& 
    1 + \frac{1}{2}\,\frac{\beta_A\alpha_B\alpha_C}
                          {(1+\alpha_A \alpha_B)(1+\alpha_A \alpha_C)},
  \label{betaABC}
\end{eqnarray}
where $A$, $B$, $C$ denote {\em a priori} three bodies, but $B=C$ is allowed. 
The relativistic periastron advance, proportional to $(2+2\gamma^\mathrm{PPN} - 
\beta^\mathrm{PPN})$ in the PPN formalism, becomes now a combination of the 
above expressions, explicitly written in eq.~(9.20a) of \cite{de92}.

But the most spectacular deviation from GR is
that scalar waves are now also emitted by any binary system, thus
contributing to the observed variation of the orbital period. For
asymmetric systems, notably the neutron star-white dwarf
binary studied in the present paper, the main contribution comes
from dipolar waves:
\begin{equation}
\label{PdotDipole}
\dot P_b^D = - 2 \pi n_b \frac{G_\ast\,M_c}{c^3}\,\frac{q}{q +1 }\,
             \frac{1 + e^2/2}{(1 - e^2)^{5/2}}\,
             (\alpha_p-\alpha_c)^2,
\end{equation}
where the eccentricity $e$ is negligible in our case.
[The lowest-order expansion of $(\alpha_p-\alpha_c)^2$ in powers of the 
sensitivities $s_{p,c}$ is denoted as $\kappa_D {\cal S}^2$ in 
eq.~(\ref{eq:pbdotD}) above. In the present section, we are 
numerically taking into account the full nonlinear dependence on the bodies' 
self-gravity.]
The companion's scalar charge $\alpha_c \approx \alpha_0$ because of its 
small binding energy, while the pulsar's scalar charge $\alpha_p$ may be of 
order 1 in some theories even if $\alpha_0\approx 0$ \citep{de93,de96b}. 
The orbital decay from dipolar gravitational 
wave emission (eq.~(\ref{PdotDipole})), which is of order
$\mathcal{O}(1/c^3)$, is thus generically much larger than the usual quadrupole 
of order $\mathcal{O}(1/c^5)$. An observed $\dot P_b$ consistent with general 
relativity therefore strongly constrains scalar-tensor theories.

This is illustrated in Fig.~\ref{ScalarTensorPlane}, where we also display the
1-$\sigma$ constraints imposed by Solar-System tests (the Cassini measurement of the
Shapiro delay, see Bertotti, Iess \& Tortora 2003 and the aforementioned
LLR experiment, Hofmann, M\"uller \& Biskupek 2010\nocite{bit03,hmb10})
and binary pulsars, according to the latest literature (Weisberg, Nice \&
Taylor 2010 for PSR~B1913+16, Stairs et al.~2002 for PSR~B1534+12, Kramer et 
al.~2006 for PSR~J0737$-$3039A/B, Bhat, Bailes \& Verbiest 2008 for 
PSR~J1141$-$6545 and Gonzalez et al.~2011 for the SEP test with an ensemble of 
binary pulsars with wide orbits and low 
eccentricities)\nocite{wnt10,sttw02,ksm+06,bbv08,gsf+11}.
For PSRs~B1913+16 and B1534+12, we multiplied the error bars on their measured $
\dot P_b^\mathrm{Int}$ by two because of the known uncertainties on their 
distances, which preclude an accurate estimate of the kinematic contributions to 
their orbital decay.

Note that the LLR constraints present a (deformed) vertical asymptote at 
$\beta_0 = -1$, whereas SEP and dipolar-radiation dominated constraints (from 
PSRs J1738+0333 and J1141$-$6545) exhibit one for $\beta_0 \approx -1.5$. This 
difference comes from the higher-order corrections in powers of the 
sensitivities, that we neglected in eq.~(\ref{eq:pbdotD}) above but which are 
taken into account in the present section. These higher-order terms are 
vanishingly small in the Earth-Moon system relevant to LLR, but they are 
numerically significant for neutron stars, and can even dominate the 
lowest-order contributions. This is notably the case when $\beta_0 = 
-1-\alpha_0^2$, which implies $\kappa_D = 0$ and a dipolar radiation actually 
starting at order ${\cal O}(s^4)$ instead of ${\cal O}(s^2)$. However, the 
proximity of these two vertical asymptotes illustrates that Fig.~7 would keep a 
similar shape even if we neglected all higher-order corrections. This would just 
slightly shift the various curves, as would also do a different choice of the 
nuclear EOS. This justifies a posteriori the lowest-order truncation used in
our analysis of Section~\ref{sec:generic_tests} above, even for
non-negligible values of $|\beta_0|$. For large values of this
coupling constant, the higher-order terms are responsible for the
dissymmetry of Fig.~7 with respect to the sign of $1+\beta_0$.

A comment on the companion mass used in eq.~(\ref{PdotDipole}). The mass of the 
white dwarf companion is derived from the optical data of Paper I, which yields 
$GM_c$, and not $G_\ast M_c$. The difference is a neglibible
factor $1 + \alpha_0^2$, which we anyway take into account when calculating
the constraints on the matter-scalar coupling constants $\alpha_0$ and 
$\beta_0$.

Figure~\ref{ScalarTensorPlane} shows that scalar-tensor theories with a 
quadratic matter-scalar coupling $\beta_0 < -5$ are forbidden, whatever the
value of the linear coupling $\alpha_0$. This is due to the nonperturbative 
strong-field effects studied in \cite{de93,de96b}.
For $\beta_0 > -5$, the limits on $\alpha_0$ are now derived 
either from the Cassini experiment or from \psr. For
positive $\beta_0$ Solar System tests used to provide the best
constraints on $\alpha_0$, but this has recently changed:
PSR J1141$-$6545 \citep{bbv08} is more constraining than the Solar System tests 
for $\beta \gtrsim 7$ and \psr\ is now the most constraining of all for
$\beta_0 > 0.7$. The same is true for
the $-4.8 < \beta_0 < -2.4$ range. The special case $\beta_0 = 0$
(the  Jordan-Fierz-Brans-Dicke theory of gravity) is in the region
where the Cassini  experiment is still more sensitive.
Our 1-$\sigma$ pulsar limit $\alpha_0^2 < 2 \times 10^{-5}$
converts into $\omega_{\rm BD} > 25000$. This is within a factor of 1.7 of
the precision of the Cassini experiment. We obtain the same constraint in the 
massive Brans-Dicke theories recently considered in \cite{abwz12}
when the scalar's mass $m_s c^2 < h/P_b = 1.35 \times 10^{-19} \, \rm eV$
(where $h$ is Planck's constant), and no longer any significant constraint for 
larger scalar masses, consistently with Fig.~1 of that reference.

Overall, \psr\ provides significantly better constraints 
than the previous best binary pulsar experiment, PSR~J1141$-$6545
(Fig.~\ref{ScalarTensorPlane}).
If the limits obtained with that or other systems improve in the
near future that would represent an important confirmation of 
the results obtained in this paper.

%%%%%%%%%%%%%%%%%%%%%%%%%%%%%%%%%%%%%%%%%%%%%%%%%%%%%%%%%%%%%%%%%%%%%%%%%%%%%%%%

\section{Constraints on $\mathrm{TeVeS}$-like theories}
\label{sec:TeVeS}

\begin{figure}
\includegraphics[width=8.5cm]{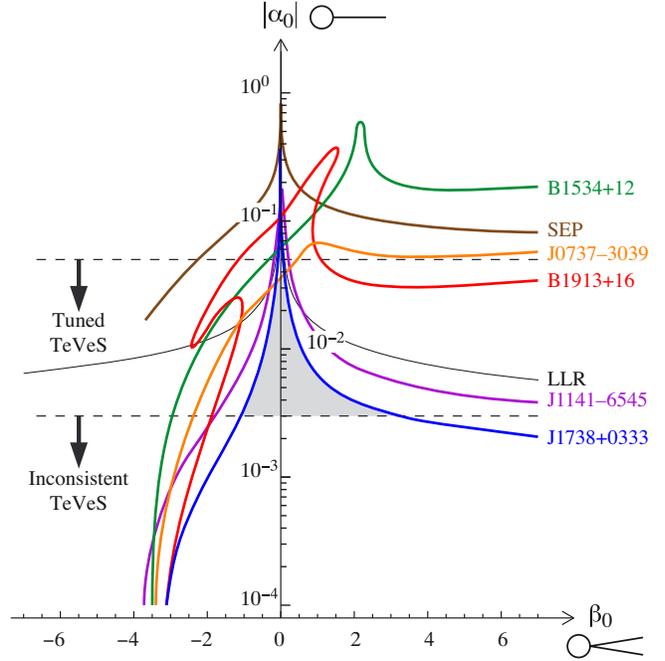}
\caption{\label{TeVeSPlane}Similar theory plane as in
Fig.~\ref{ScalarTensorPlane}, but now for the (non-conformal)
matter-scalar coupling described in the text, generalizing the
TeVeS model. Above the upper horizontal dashed line, the
nonlinear kinetic term of the scalar field may be a natural
function; between the two dashed lines, this function needs to be
tuned; and below the lower dashed line, it cannot exist any
longer. The allowed region is shaded. It excludes general
relativity ($\alpha_0 = \beta_0 = 0$) because such models are
built to predict modified Newtonian dynamics (MOND) at large
distances. Note that binary pulsars are more constraining than
Solar-System tests for this class of models (and that the Cassini
bound of Fig.~\ref{ScalarTensorPlane} does not exist any longer
here). For a generic nonzero $\beta_0$, \psr\ is again
the most constraining binary pulsar, while for $\beta_0 \approx
0$, the magnitude of $|\alpha_0|$ is bounded by the
J0737$-$3039 system.}
\end{figure}

A tensor-vector-scalar (TeVeS) theory of gravity has been proposed by 
\cite{bek04} to account for galaxy rotation curves and weak lensing 
without the need for dark matter. This is a relativistic realization of the 
modified Newtonian dynamics (MOND) proposal \citep{mil83}, which introduces a 
fundamental acceleration scale $a_0 \approx 10^{-10}\,\rm m\,s^{-2}$ (not 
to be confused with the matter-scalar coupling constant $\alpha_0$ defined 
above). One of the difficulties is to be able to predict significant deviations 
from Newtonian gravity at large distances, while being consistent with
Solar-System and binary-pulsar tests of GR at smaller scales
\citep{san96,be07}. Indeed, the scalar-field kinetic term of TeVeS is an unknown 
nonlinear function, which must take different forms at small and large 
distances. This function can have a natural shape only if $|\alpha_0| > 
\sqrt{r_{\odot U}/r_{\rm MOND}} \approx 0.05$, where $r_{\odot U}$ is the 
orbital radius of Uranus and $r_{\rm MOND} = \sqrt{GM_\odot/a_0} 
\approx 7000\,{\rm AU} \approx 0.1\,$lt-yr, otherwise the model would 
predict anomalies  too large to be consistent with planetary ephemerides
\citep{lfm+09}. Below this value, the function needs to be tuned,
and even fine-tuned for much smaller $|\alpha_0|$, and it merely
cannot exist any longer if $|\alpha_0| < r_{\odot U}/r_{\rm MOND} \approx 0.003$
(it would need to be bi-valued). Using binary pulsars to constrain the
matter-scalar coupling constant $\alpha_0$ within TeVeS is thus particularly
interesting.

In the following, we shall not take into account all the subtle
details of TeVeS, which are not relevant for our conclusions. In
particular, Bekenstein introduced specific scalar-vector
couplings to avoid superluminal propagation, but this is actually
not necessary to respect causality \citep{bru07,be07,bmv08}.
We will thus focus on the small-distance behavior
of this theory, and assume that the scalar-field kinetic term
takes its standard form in this limit. We will also neglect the
contributions of the vector field to the dynamics and the
gravitational radiation, because they depend again on some
coupling constant which can be chosen small enough. Let us just
underline that if the vector field carries away some significant
amount of energy, then binary-pulsar data are even more
constraining than what we obtain below. Neglecting these
contributions is thus a conservative choice. To simplify, we will
here assume that the vector field of TeVeS is aligned with the
proper time direction of matter.

The crucial difference of TeVeS, with respect to the standard
scalar-tensor theories of Section~\ref{sec:STe}, is the expression of
the physical metric $\tilde g_{\mu\nu}$ to which matter is
universally coupled. In the rest frame of matter, it reads
$\tilde g_{00} = A^2(\varphi) g_{00}$ for the time component, but
$\tilde g_{ij} = A^{-2}(\varphi) g_{ij}$ for the spatial ones.
The actual model constructed in \cite{bek04} assumes $A(\varphi)
= \exp (\alpha_0 \varphi)$, but we shall here generalize it and
write $\ln A(\varphi) = \ln A(\varphi_0) + \alpha_0
(\varphi-\varphi_0) + \frac{1}{2} \beta_0 (\varphi-\varphi_0)^2$,
as in Section~\ref{sec:STe} above. We are thus considering here a class
of TeVeS-like theories rather than the single model of \cite{bek04}.

This specific form of the physical metric results in several
important differences with respect to Section~\ref{sec:STe}. First of
all, to compute the matter-scalar coupling parameters $\alpha_A$
and $\beta_A$ corresponding to a strongly self-gravitating body
$A$, one must numerically integrate new field equations. To save
space, we just quote here the differences with respect to \cite{de96b}, 
where the standard scalar-tensor case was
discussed in detail: all factors $A^4(\varphi)$ in its
Eqs.~(3.6a,b,d,e,h) and (3.14d,h) should be replaced by
$A^{-2}(\varphi)$, while the factor $A^3(\varphi)$ of its
eq.~(3.6f) should become $A^{-3}(\varphi)$. Finally, the
$(\tilde\varepsilon-3\tilde p)$ term of its eq.~(3.6d) and
(3.14d) should be replaced by the sum $(\tilde\varepsilon
+3\tilde p)$. [On the other hand, note that the
$(\tilde\varepsilon -\tilde p)$ term of its eq.~(3.6d) does not
change, nor the sign of $\alpha(\varphi)$ in its eq.~(3.6e).]
These modifications are consistent with the results of \cite{lsg08}. 
After numerically integrating these field equations, starting from
central values of the scalar field and matter density, one can
extract the mass of the neutron star and its effective scalar
charge $\alpha_A$ in the same way as in \cite{de96b}. The
effective quadratic coupling $\beta_A$ is then defined as
$\partial \alpha_A/\partial \varphi_0$ for a fixed baryonic mass.
For some binary-pulsar tests, it is also necessary to compute the
star's moment of inertia, and its derivative with respect to
the background scalar field $\varphi_0$ at constant baryonic mass.

The form of the physical metric also changes the post-Newtonian predictions of 
the theory. Instead of eqs.~(\ref{GAB}) and (\ref{gammaAB}) above, we now get
\begin{eqnarray}
  \tilde G_{AB} &=& G_\ast (1+\alpha_A\alpha_B),
  \label{GABTeVeS}\\
  \gamma_{AB} &=& \gamma^\mathrm{PPN} = 1,
  \label{gammaABTeVeS}
\end{eqnarray}
while eq.~(\ref{betaABC}) is unchanged. Note that the effective gravitational 
constant $\tilde G_{AB}$ does not depend any longer on $A^2(\varphi_0)$. In the 
actual TeVeS theory of \cite{bek04}, where $\beta_0 = 0$, the 
weak-field effective gravitational constant is thus a true constant given by 
$G_\ast (1+\alpha_0^2)$, in particular time-independent as was first derived in 
\cite{bs08}. This constancy also explains why LLR and other SEP tests do not 
constrain at all the theories with $\beta_0 = 0$, as illustrated in 
Fig.~\ref{TeVeSPlane}.

The expression for the periastron advance takes the same form as
in standard scalar-tensor theories in terms of $\gamma_{AB}$ and
$\beta^A_{BC}$, but now with $\gamma_{AB} = 1$. In the weak-field
conditions of the Solar System, the fact that $\gamma^\mathrm{PPN}
= 1$, like in GR, explains why the Cassini
time-delay experiment does not provide any constraint on
TeVeS-like theories.

All the other predictions of scalar-tensor theories, including
the dipolar damping given by eq.~(\ref{PdotDipole}), keep the same forms 
in terms of the body-dependent quantities $\alpha_A$ and $\beta_A$,
but their numerical values do differ because of the numerical
integrations described above. For $\beta_0 = 0$, i.e., the actual
TeVeS theory of Bekenstein (2004), one finds that $\alpha_A =
\alpha_0$ independently of the body. This can be deduced from
eq.~(4.14) of \cite{de96a}, where the scalar field is no longer
sourced by the trace of the energy-momentum tensor, as in
standard scalar-tensor theories, but by the same combination
($\tilde\varepsilon + 3\tilde p$) as the gravitational potential.
The constancy of $\alpha_A = \alpha_0$, when $\beta_0 = 0$, is
also confirmed by our numerical integrations. The consequence of
this extra symmetry of TeVeS is that the dipolar radiation
predicted by eq.~(\ref{PdotDipole}) merely vanishes, and asymmetrical 
neutron star-white dwarf binaries like \psr\ no longer provide any
stringent constraints. This is illustrated in
Fig.~\ref{TeVeSPlane}, where several binary-pulsar tests now present
a bump on the vertical axis, $\beta_0 = 0$. As compared to
Fig.~\ref{ScalarTensorPlane}, the main effect of the
non-conformal matter-scalar coupling assumed in TeVeS-like models
is thus to displace these bumps from $\beta_0 \sim -1$ or $-2$ to
the axis $\beta_0 = 0$.

For generic TeVeS-like models with $|\beta_0| > 0.1$, we find
again that \psr\ is the most constraining binary pulsar.
We also note that several binary pulsar tests are now
{\em more} constraining than all Solar-System experiments
(i.e., the LLR line of Fig.~\ref{TeVeSPlane}, while the
perihelion shift of Mercury \citep{sha90b} gives even weaker
constraints, and all deflection or time-delay predictions
coincide with those of GR at the first
post-Newtonian order).

From a theoretical point of view, the most natural value of the
linear matter-scalar coupling constant $|\alpha_0|$ would be
of order unity.
Figure~\ref{TeVeSPlane} shows this is experimentally forbidden,
even on the vertical axis $\beta_0 = 0$ thanks to several binary
pulsars whose tests do not rely on the magnitude of the dipolar
radiation\footnote{The two ``horns'' of the constraints imposed by PSR
B1913+16 come from the fact that this system provides only three
post-Keplerian observables, i.e., only one test, and some
fine-tuned models can thus pass it.}. They actually impose that
$|\alpha_0| < 0.035 < \sqrt{r_{\odot U}/r_{\rm MOND}}$, therefore
TeVeS needs to assume an unnatural shape of the function defining
its scalar's dynamics.

A possible way to avoid fine tuning in TeVeS-like models has
been proposed in \cite{bde11}, where it was shown that
$|\alpha_0| = 1$ can be consistent with Solar-System and
binary-pulsar tests, thanks to a screening of all scalar-field
effects at small distances. Therefore, the results of the
present paper do not rule out TeVeS, but show that its original
2004 formulation by Bekenstein may need to be amended. At present,
even its original writing is consistent, although it does need
some tuning.

%%%%%%%%%%%%%%%%%%%%%%%%%%%%%%%%%%%%%%%%%%%%%%%%%%%%%%%%%%%%%%%%%%%%%%%%%%%%%%%%

\section{Conclusions and Prospects}
\label{sec:conclusions}

It is quite fortunate that the timing precision of \psr\ allows a precise 
measurement of the key observables necessary for an estimation of the 
intrinsic orbital decay ($\dot{P}_b$, $\mu_{\alpha}$, $\mu_{\delta}$ and $\pi_x
$) and that the optical observations provide a precise estimate of a general
relativistic prediction for the orbital decay, thus allowing a stringent limit 
on the emission rate for dipolar GWs. As a result of this,
\psr\ is already the most constraining binary pulsar for (conformally-coupled) 
scalar-tensor theories of gravity.

\psr\ is also the most constraining test of TeVeS-like theories when the 
quadratic matter-scalar coupling constant $|\beta_0| \ga 0.1$. In fact, for
$\beta_0 < -1$ and $\beta_0 > 3$, such theories are excluded altogether. 
Bekenstein's TeVeS (a special case with $\beta_0 = 0$) is still allowed by the 
results of this experiment, but already needs some tuning given the small 
limit $|\alpha_0| < 0.035$ that we obtain from the double pulsar results 
\citep{ksm+06}. We note that the precision of the latter result has greatly 
improved since 2006 and will be presented in a forthcoming publication 
(Kramer et al., in prep.). This will significantly reduce the allowed values of 
$|\alpha_0|$ in the gap around $\beta_0=0$. As a consequence, all surviving 
TeVeS-like theories will have to be unnaturally fine-tuned, including 
Bekenstein's TeVeS.

Continued timing of \psr\ will give us a much more precise measurement of
$\dot{P}_b$ and (to a smaller extent) of $\pi_x$. This will allow a
further improvement of the 
estimate of the intrinsic orbital decay. However, unless we can make a more 
precise determination of $m_c$, we will not be able to improve upon the current 
estimate of $\dot{P}_b^{\rm GR}$; this implies that the precision
of this test is 
unlikely to improve by more than a factor of 2. Despite that, a precise 
measurement of $\dot{P}_b^{\rm Int}$ is still very useful: together with the 
measurement of $q$, it will eventually allow (with the assumption of the 
correctness of GR) a very precise estimate of the system masses 
and a precise calibration of the mass-radius relation for white dwarfs.

%%%%%%%%%%%%%%%%%%%%%%%%%%%%%%%%%%%%%%%%%%%%%%%%%%%%%%%%%%%%%%%%%%%%%%%%%%%%%%%%

\section*{Acknowledgments}

We are grateful to Julia Deneva for helping with many observations, to Arun 
Venkataraman for help managing the large data sets resulting from this project 
year after year, Hector Hern{\'a}ndez and the Arecibo operators for helping us 
optimize scheduling and usage of telescope time and Marten van Kerkwijk for 
useful comments and suggestions. We also thank Paul Demorest, Rob Ferdman, David 
Nice and Don Backer for permission to use the Arecibo Signal Processor to check 
the accuracy of our timing. P.F. and J.P.W.V. gratefully acknowledge the
financial support 
by the European Research Council for the ERC Starting Grant BEACON under 
contract no. 279702. G.E.-F. wishes to thank the Max-Planck-Institut f\"ur 
Radioastronomie for its kind hospitality during part of this work. He was also 
in part supported by the ANR grant ``THALES''. During most
of this work J.P.W.V. was supported by the 
European Union  under Marie Curie Intra-European Fellowship 236394. Pulsar 
research at UBC is supported by an NSERC Discovery Grant.
 Arecibo Observatory's William Gordon Telescope  is a facility of the 
NSF. During this work it was operated through a cooperative agreement with 
Cornell University; it is now operated by SRI/USRA and Universidad 
Metropolitana. The Parkes Observatory is part of the Australia Telescope which 
is funded by the Commonwealth of Australia for operation as a National Facility 
managed by CSIRO. The Westerbork Synthesis Radio Telescope is operated by ASTRON 
(Netherlands Foundation for Research in Astronomy) with support from The 
Netherlands Foundation for Scientific Research NWO. We would like to thank the 
referee, Thibault Damour, for constructive and detailed comments, which
helped to improve the manuscript.

%%%%%%%%%%%%%%%%%%%%%%%%%%%%%%%%%%%%%%%%%%%%%%%%%%%%%%%%%%%%%%%%%%%%%%%%%%%%%%%%

%%%%%%%%%%%%%%%%%%%%%%%%%%%%%%%%%%%%%%%%%%%%%%%%%%%%%%%%%%%%%%%%%%%%%%%%%%%%%%%%

\appendix
\section{Constraining the mass loss of the WD companion}
\label{sec:mass_loss}

\begin{figure}
  \includegraphics[width=10cm]{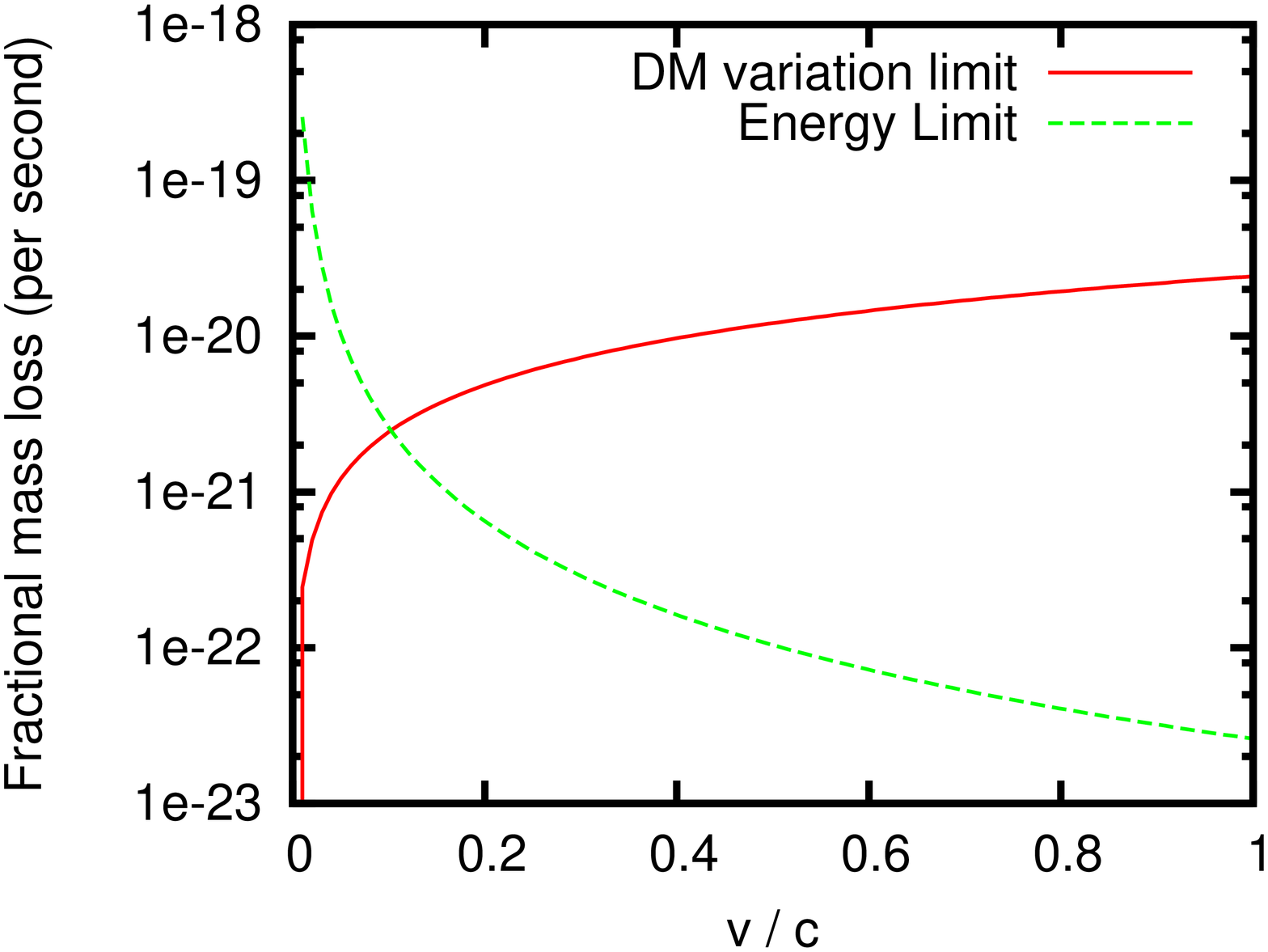}
  \caption{Upper limits for mass loss from the white dwarf derived
  from upper limit of pulsar spin-down
  energy incident on the white dwarf and 1-$\sigma$ upper limit
  on $\Delta \rm DM$. Their combination excludes a fractional
  mass loss larger than $2.5 \times 10^{-21}\,\rm s^{-1}$.}
  \label{fig:mass_loss}
\end{figure}

The mass loss of the companion, $\dot{M}_c$, can be constrained
by assuming (safely) that the energy driving the process
comes from the pulsar. Its rotational energy loss is given by
$\dot{E} = - I_p \Omega_p \dot{\Omega}_p
= 4 \pi^2 I_p \dot{P}_{\rm Int} P^{-3} = 4.4 
\times 10^{33}\,\rm erg \,  s^{-1}$,
where $\dot{P}_{\rm Int}$ is its intrinsic spin-down
(which is calculated from the observed spin-down after taking into
account kinematic effects, see Section~\ref{sec:intrinsic_decay}),
$I_p\sim 10^{45}$ g\,cm$^{2}$ its moment of inertia and
$\Omega_p$ its angular velocity.
Some of this radiated energy impacts the white dwarf;
its maximum occurs when the pulsar is beaming all the power right on the
orbital plane, in which case the white dwarf intercepts a fraction
given by $f = 2 R_c / (2 \pi a)$, where $R_c$ is the
WD companion radius, $0.037\, \rm R_{\odot}$ (see Paper I) and
$a = xc(q+1)/\sin i$ is the orbital separation. This then
provides the escaping particles with a velocity $v_0$ such that,
through conservation of energy
\begin{equation}
\label{eq:energy}
\frac{1}{2} \dot{M}_c v_0^2 < \dot{E} \, \frac{R_c}{\pi a} \Rightarrow
\dot{M}_c < 2.60 \times 10^{-23} \, {\rm s^{-1}}
\left( \frac{v_0}{c} \right)^{-2}\, M_t,
\end{equation}
where we express the mass loss rate as a fraction of the total
mass of the system.
To first approximation, this initial velocity has to be larger
than the escape velocity from the surface of the WD,
$v_e = \sqrt{2 G M_c/R_c} = 1.37 \times 10^{3}\, {\rm km \, s^{-1}} = 0.0046\, 
c$; this immediately yields $\dot{M_c}/M_t < 1.2 \times 10^{-18}\, \rm s^{-1}$.

In what follows, we assume that the gas escapes from the WD
equally in all directions. The matter density is then given by
$\rho(r) = \dot{M}_c / [4 \pi r^2 v(r)]$ and the electron density by
\begin{equation}
n(r) = \frac{\dot{M}_c}{4 \pi r^2 X \mu_P v(r)},
\end{equation}
where $r$ is the distance to the center of the WD,
$\mu_P$ is the proton mass and $X$ is the number of nucleons per
electron (1 for Hydrogen, 2 for heavy elements). Therefore, for
any particular density profile $n(r)$, the larger the plasma velocity,
the larger is $\dot{M}_c$. As calculated below, for the latter to
be of any practical significance in the present case, $v_{0}$
has to be at least of the order of a few percent of the speed of
light $c$. In this case the plasma would be
little affected by the WD escape velocity and have
a nearly constant velocity.
This results in an upper limit for the density profile given by
$n(r) = \dot{M}_c / (4 \pi r^2 X \mu_P v_0)$. Therefore, the
upper limit for the extra column density along the length $l$ of the
line of sight from the pulsar to Earth is given by
\begin{equation}\label{eq:I}
  H = \int_{-r_p\cos\psi}^{\infty} n(r) \, dl
    = \frac{\dot{M}_c}{4 \pi X \mu_P v_0} \, \frac{\psi}{r_p \sin\psi} , 
\end{equation}
where $r_p$ is the distance between the pulsar and the WD, and $\psi$ is
the angle between the direction to the pulsar and the direction to the observer 
as seen from the WD, in such a way that $r = [(r_p\sin\psi)^2 + l^2]^{1/2}$. 
This equation is similar to eq. (5) of \cite{ktm96}.

For a very low-eccentricity orbit $r_p \simeq a$.
For superior and inferior conjunction $\psi = \pi/2 + i$ and $\psi = \pi/2 - i$,
respectively, therefore
\begin{equation}
  \Delta H \equiv H_{\rm sup} - H_{\rm inf}  
   = \frac{1}{2\pi} \, \frac{\dot{M}_c}{ X \mu_P v_0} \, \frac{i}{a \sin i} . 
\end{equation}
Dividing by the total mass of the binary $M_t = M_p + M_c$ we obtain
\begin{eqnarray}
\frac{\dot{M}_c}{M_t} 
  &=& 2 \pi v_0 X \frac{\mu_P}{M_c} \, 
      \frac{xc\cot i}{i} \, \Delta H \nonumber\\
  &=& 1.97 \times 10^{-16} {\rm s^{-1}} \,
      \left(\frac{v_0}{c}\right) {\rm \Delta DM} \label{eq:dm}.
\end{eqnarray}
In this expression we use
$\Delta H = \Delta \rm DM \, \times 3.0857\, \times 10^{18} \,cm\,pc^{-1}$
to convert the column density to a DM,
$X = 2$ as an upper limit for $X$ and the $x$ and $i$ measured
for this system (see Table~\ref{tab:parameters}).
The DM variations for \psr\ are displayed in the left plot of
Fig.~\ref{fig:orbit}. Fitting an equivalent of eq.~(\ref{eq:I}) to
these DMs (also displayed in the same plot)
we obtain $\Delta \rm DM = (0.24 \pm 1.09) \times 10^{-4}\,\rm cm^{-3}\,pc$,
which is basically consistent with no observed dispersive delays.

Using the upper limit represented by eq.~(\ref{eq:energy}) and 
the 1-$\sigma$ upper limit (84\% C. L.)
for $\Delta \rm DM$ in eq.~(\ref{eq:dm}) 
we obtain $\frac{\dot{M}_c}{M_t} < 2.5 \times 10^{-21}\,\rm s^{-1}$,
which happens at a velocity of about $0.1\, c$
(see Fig.~\ref{fig:mass_loss}). This is large enough for the velocity
to be nearly constant as a function of $r$ (eq.~\ref{eq:dm}) but small
enough for the Newtonian expression for the kinetic energy 
(eq.~\ref{eq:energy}) to provide a good approximation.

It is possible that the gas is being ejected slowly compared
to the pulsar wind, in which
case it might be blown away from the pulsar by its wind, forming a
highly anysotropic ``cometary" tail. In that case the
low orbital inclination of this system would mean that the plasma
emanating from the WD might avoid crossing the line of sight,
producing no extra dispersive delays detectable at the Earth.
In such a case we can only rely on the energy balance to
constrain the mass loss, which at these low plasma velocities
allows, in the extreme case discussed above, for very substantial mass loss.
There are, however, other binary pulsars with companion WDs of
similar mass
like PSR~J0751+1807 \citep{nss+05} and PSR~J1909$-$3744
\citep{jhb+05} that happen to be observed at much higher inclinations,
and in no case is such a cometary tail ever observed.

%----------


\begin{thebibliography}{}

\bibitem[\protect\citeauthoryear {Alsing et al.}{2011}]{abwz12}
{Alsing} J., {Berti} E., {Will} C., {Zaglauer} H., 2012, 
Phys.~Rev., D 85, 064041

\bibitem[\protect\citeauthoryear {Antoniadis et al.}{2012}]{akk+12}
{Antoniadis} J., {van Kerkwijk} M.~H., {Koester} D., {Freire} P.~C.~C.,
{Wex} N., {Tauris} T.~M., {Kramer} M., 2012, 
MNRAS, in press (arXiv:1204.3948)

\bibitem[\protect\citeauthoryear {Babichev, Deffayet \& 
Esposito-Far\`ese}{2011}]{bde11}
{Babichev} E., {Deffayet} C., {Esposito-Far\`ese} G., 
2011, Phys.~Rev., D 84, 061502

\bibitem[\protect\citeauthoryear {Babichev, Mukhanov \& Vikman}{2008}]{bmv08}
{Babichev} E., {Mukhanov} V., {Vikman} A., 2008, JHEP, 0802, 101

\bibitem[\protect\citeauthoryear {Bassa et al.}{2006}]{bkkv06}
{Bassa} C.~G., {van Kerkwijk} M.~H., {Koester} D., {Verbunt} F., 
2006, A\&A, 456, 295

\bibitem[\protect\citeauthoryear {Bekenstein}{2004}]{bek04}
{Bekenstein} J.~D., 
2004, Phys.~Rev., D 70, 083509.
[Erratum-ibid., D 71, 069901 (2005)]

\bibitem[\protect\citeauthoryear {Bekenstein \& Sagi}{2008}]{bs08}
{Bekenstein} J.~D.,  {Sagi} E., 
2008, Phys.~Rev., D 77, 103512

\bibitem[\protect\citeauthoryear {Bertotti, Iess \& Tortora}{2003}]{bit03}
{Bertotti} B., {Iess} L., {Tortora} P., 
2003, Nature, 425, 374

\bibitem[\protect\citeauthoryear {Bhat, Bailes \& Verbiest}{2008}]{bbv08}
{Bhat} N. D.~R., {Bailes} M., {Verbiest} J.~P.~W., 
2008, Phys.~Rev., D 77, 124017

\bibitem[\protect\citeauthoryear {Brans \& Dicke}{1961}]{bd61}
{Brans} C.,  {Dicke} R.~H., 
1961, Phys. Rev., 124, 925

\bibitem[\protect\citeauthoryear {Bruneton}{2007}]{bru07}
{Bruneton} J.-P., 
2007, Phys.~Rev., D 75, 085013

\bibitem[\protect\citeauthoryear {Bruneton \& Esposito-Far\`ese}{2007}]{be07}
{Bruneton} J.-P.,  {Esposito-Far\`ese} G., 
2007, Phys.~Rev., D 76, 124012

\bibitem[\protect\citeauthoryear {Cordes \& Lazio}{2001}]{cl01}
{Cordes} J.~M.,  {Lazio} T.~J.~W., 
2001, ApJ, 549, 997

\bibitem[\protect\citeauthoryear {Damour}{2007}]{dam07}
{Damour} T., 2007, 
lecture ``Binary Systems as Test-beds of Gravity Theories'', 
arXiv:0704.0749

\bibitem[\protect\citeauthoryear {Damour \& Deruelle}{1985}]{dd85}
{Damour}, T., Deruelle, N.\ 1985, Ann.~Inst.~Henri 
Poincar{\'e},  
Phys.~Th{\'e}or., 43, 107.

\bibitem[\protect\citeauthoryear {Damour \& Deruelle}{1986}]{dd86}
{Damour}, T., Deruelle, N.\ 1986, Ann.~Inst.~Henri Poincar{\'e}, 
Phys.~Th{\'e}or., 44, 263.

\bibitem[\protect\citeauthoryear {Damour \& Esposito-Far\`ese}{1992}]{de92}
{Damour} T.,  {Esposito-Far\`ese} G., 
1992, Class.~Quantum~Grav., 9, 2093

\bibitem[\protect\citeauthoryear {Damour \& Esposito-Far\`ese}{1993}]{de93}
{Damour} T.,  {Esposito-Far\`ese} G., 
1993, Phys.~Rev.~Lett., 70, 2220

\bibitem[\protect\citeauthoryear {Damour \& Esposito-Far\`ese}{1996a}]{de96a}
{Damour} T.,  {Esposito-Far\`ese} G., 
1996a, Phys.~Rev., D 53, 5541

\bibitem[\protect\citeauthoryear {Damour \& Esposito-Far\`ese}{1996b}]{de96b}
{Damour} T.,  {Esposito-Far\`ese} G., 
1996b, Phys.~Rev., D 54, 1474

\bibitem[\protect\citeauthoryear {Damour \& Esposito-Far\`ese}{1998}]{de98}
{Damour} T.,  {Esposito-Far\`ese} G., 
1998, Phys.~Rev., D 58, 042001

\bibitem[\protect\citeauthoryear {Damour, Gibbons \& Taylor}{1988}]{dgt88}
{Damour} T., {Gibbons} G.~W., {Taylor} J.~H., 
1988, Phys.~Rev.~Lett., 61, 1151

\bibitem[\protect\citeauthoryear {Damour \& Taylor}{1991}]{dt91}
{Damour} T.,  {Taylor} J.~H., 
1991, ApJ, 366, 501

\bibitem[\protect\citeauthoryear {Deller et al.}{2008}]{dvtb08}
{Deller} A.~T., {Verbiest} J.~P.~W., {Tingay} S.~J., {Bailes} M., 
2008, ApJ, 685, L67

\bibitem[\protect\citeauthoryear {Demorest}{2007}]{dem07}
{Demorest} P.~B., 
2007, PhD thesis, University of California, Berkeley

\bibitem[\protect\citeauthoryear {Dowd, Sisk \& Hagen}{2000}]{dsh00}
{Dowd} A., {Sisk} W., {Hagen} J., 
2000, in ``Pulsar Astronomy --- 2000 and Beyond,
IAU Colloquium 177'', ed.\ M.~Kramer, N.~Wex \& R.~Wielebinski, 
(San Francisco: Astronomical Society of the Pacific), p.~275

\bibitem[\protect\citeauthoryear {Eardley}{1975}]{ear75}
{Eardley} D.~M., 1975, ApJ, 196, L59.

\bibitem[\protect\citeauthoryear {Eddington}{1923}]{edd23}
{Eddington} A., 1923, ``The Mathematical Theory of Relativity'', 
(London: Cambridge University Press)

\bibitem[\protect\citeauthoryear {Edwards, Hobbs \& Manchester}{2006}]{ehm06}
{Edwards} R.~T., {Hobbs} G.~B., {Manchester} R.~N., 2006, MNRAS, 372, 1549

\bibitem[\protect\citeauthoryear {Fierz}{1956}]{fie56}
{Fierz} M., 1956, Helv.~Phys.~Acta, 29, 128

\bibitem[\protect\citeauthoryear {Folkner, Williams \& Boggs}{2008}]{fwb08}
{Folkner} W.~M., {Williams} J.~G., {Boggs} D.~H., 2008, 
``The planetary and Lunar ephemeris DE 421'', 
JPL Memorandum IOM 343R-08-003

\bibitem[\protect\citeauthoryear {Freire \& Wex}{2010}]{fw10}
{Freire} P.~C.~C.,  {Wex} N., 2010, MNRAS, 409, 199

\bibitem[\protect\citeauthoryear {Gonzalez et al.}{2011}]{gsf+11}
{Gonzalez} M.~E., {Stairs} I.~H., {Ferdman} R.~D., {Freire} P.~C.~C., 
{Nice} D.~J., {Demorest} P.~B., {Ransom} S.~M., {Kramer} M., {Camilo} F., 
{Hobbs} G., {Manchester} R.~N., {Lyne} A. G., 2011, 
ApJ, 743, 102

\bibitem[\protect\citeauthoryear {Hobbs, Edwards \& Manchester}{2006}]{hem06}
{Hobbs} G.~B., {Edwards} R.~T., {Manchester} R.~N., 2006, 
MNRAS, 369, 655

\bibitem[\protect\citeauthoryear {Hofmann, M\"uller \& Biskupek}{2010}]{hmb10}
{Hofmann} F., {M\"uller} J., {Biskupek} L., 2010, A\&A, 522, L5

\bibitem[\protect\citeauthoryear {Jacoby}{2005}]{jac05}
{Jacoby} B.~A., 2005,
PhD thesis, California Institute of Technology, California, USA

\bibitem[\protect\citeauthoryear {Jacoby et al.}{2007}]{jbo+07}
{Jacoby} B.~A., {Bailes} M., {Ord} S.~M., {Knight} H.~S., {Hotan} A.~W.,
2007, ApJ, 656, 408

\bibitem[\protect\citeauthoryear {Jacoby et al.}{2005}]{jhb+05}
{Jacoby} B.~A., {Hotan} A., {Bailes} M., {Ord} S., {Kuklarni} S.~R.,
2005, ApJ, 629, L113

\bibitem[\protect\citeauthoryear {Jordan}{1959}]{jor59}
{Jordan} P., 1959, Z.~Phys., 157, 112

\bibitem[\protect\citeauthoryear {Karuppusamy, Stappers, 
\& van Straten}{2008}]{kss08}
{Karuppusamy} R., {Stappers} B., {van Straten} W., 2008, PASP, 120, 191

\bibitem[\protect\citeauthoryear {Kaspi, Tauris \& Manchester}{1996}]{ktm96}
{Kaspi} V.~M., {Tauris} T., {Manchester} R.~N., 1996, ApJ, 459, 717

\bibitem[\protect\citeauthoryear {Kramer et al.}{2006}]{ksm+06}
{Kramer} M., {Stairs} I.~H., {Manchester} R.~N., {McLaughlin} M.~A.,
{Lyne} A.~G., {Ferdman} R.~D., {Burgay} M., {Lorimer} D.~R., 
{Possenti} A., {D'Amico} N., {Sarkissian} J.~M., {Hobbs} G.~B., 
{Reynolds} J.~E., {Freire} P.~C.~C.,  Camilo, F.,
2006, Science, 314, 97

\bibitem[\protect\citeauthoryear {Lange et al.}{2001}]{lcw+01}
{Lange} C., {Camilo} F., {Wex} N., {Kramer} M., {Backer} D., {Lyne} A., 
\& {Doroshenko} O., 
2001, MNRAS, 326, 274

\bibitem[\protect\citeauthoryear {Laskar et al.}{2009}]{lfm+09}
{Laskar} J., {Fienga} A., {Manche} H., {Kuchynka} P., 
{Le Poncin-Lafitte} C., {Gastineau} M., 
2009, American Astronomical Society, IAU Symposium 261.~Relativity in 
Fundamental Astronomy: Dynamics, Reference Frames, and Data Analysis 
27 April -- 1 May 2009, Virginia Beach, VA, USA, 6.02; 
Bulletin of the American Astronomical Society, Vol.~41, p.~881

\bibitem[\protect\citeauthoryear {Lasky, Sotani \& Giannios}{2008}]{lsg08}
{Lasky} P.~D., {Sotani} H., {Giannios} D., 2008, Phys.~Rev., D 78, 104019

\bibitem[\protect\citeauthoryear {Lazaridis et al.}{2009}]{lwj+09}
{Lazaridis} K., {Wex} N., {Jessner} A., {Kramer} M., {Stappers} B. W.,
{Janssen} G.~H., {Desvignes} G., {Purver} M.~B., {Cognard} I., 
{Theureau} G., {Lyne} A.~G., {Jordan} C.~A., {Zensus} J.~A.,
2009, MNRAS, 400, 805

\bibitem[\protect\citeauthoryear {Lorimer \& Kramer}{2005}]{lk05}
{Lorimer} D. R.,  {Kramer} M., 2005, 
``Handbook of Pulsar Astronomy'', 
(Cambridge: Cambridge University Press)

\bibitem[\protect\citeauthoryear {Milgrom}{1983}]{mil83}
{Milgrom} M., 1983, ApJ, 270, 365

\bibitem[\protect\citeauthoryear {Motz}{1952}]{motz52}
{Motz} L., 1952, ApJ, 115, 562

\bibitem[\protect\citeauthoryear {Nice et al.}{2005}]{nss+05}
{Nice} D.~J., {Splaver} E.~M., {Stairs} I.~H., {L\"ohmer} O., 
{Jessner} A., {Kramer} M., {Cordes} J.~M., 2005, ApJ, 634, 1242

\bibitem[\protect\citeauthoryear {Nice, Stairs \& Kasian}{2008}]{nsk08}
{Nice} D.~J., {Stairs} I.~H., {Kasian} L.~E. 
2008, in 40 Years of Pulsars: Millisecond Pulsars, Magnetars and More, 
ed.\ C.~Bassa, Z.~Wang, A.~Cumming \& V.~M.~Kaspi, Vol.~983 of 
American Institute of Physics Conference Series, p.~453

\bibitem[\protect\citeauthoryear {Nice \& Taylor}{1995}]{nt95}
{Nice}, D.~J., {Taylor} J.~H., 1995, ApJ, 441, 429

\bibitem[\protect\citeauthoryear {Nordtvedt}{1990}]{nor90}
{Nordtvedt} K., 
1990, Phys.~Rev.~Lett., 65, 953

\bibitem[\protect\citeauthoryear {Riess et al.}{2009}]{rmc+09}
{Riess} A.~G., {Macri} L., {Casertano} S., {Sosey} M., {Lampeitl} H., 
{Ferguson} H.~C., {Filippenko} A.~V., {Jha} S.~W., {Li} W., {Chornock} R., 
\& {Sarkar} D., 2009, ApJ, 699, 539

\bibitem[\protect\citeauthoryear {Sanders}{1997}]{san96}
{Sanders} R.~H., 1997, ApJ, 480, 492

\bibitem[\protect\citeauthoryear {Shapiro}{1990}]{sha90b}
{Shapiro} I.~I., 1990, 
in Proceedings of the 12th International Conference on General Relativity 
and Gravitation, ed.\ N.~Ashby, D.~F.~Bartlett \& W.~Wyss, 
(Cambridge: Cambridge University Press), p.~313

\bibitem[\protect\citeauthoryear {Shklovskii}{1970}]{shk70}
{Shklovskii} I.~S., 
1970, Sov.~Astron., 13, 562

\bibitem[\protect\citeauthoryear {Smarr \& Blandford}{1976}]{sb76}
{Smarr} L.~L., {Blandford} R., 
1976, ApJ, 207, 574

\bibitem[\protect\citeauthoryear {Stairs et al.}{2002}]{sttw02}
{Stairs} I.~H., {Thorsett} S.~E., {Taylor} J.~H., {Wolszczan} A., 
2002, ApJ, 581, 501

\bibitem[\protect\citeauthoryear {Taylor}{1992}]{tay92}
{Taylor} J.~H., 1992, 
Philosophical Transactions of the Royal Society of London, A 341, 117

\bibitem[\protect\citeauthoryear {Verbiest et al.}{2009}]{vbc+09}
{Verbiest} J.~P.~W., {Bailes} M., {Coles} W.~A., {Hobbs} G.~B., 
{van Straten} W., {Champion} D.~J., {Jenet} F.~A., {Manchester} R.~N., 
{Bhat} N.~D.~R., {Sarkissian} J.~M., {Yardley} D., {Burke-Spolaor} S.,
{Hotan} A.~W.,  You, X.~P.,
2009, MNRAS, 400, 951

\bibitem[\protect\citeauthoryear {Verbiest et al.}{2008}]{vbs+08}
{Verbiest} J.~P.~W., {Bailes} M., {van Straten} W., {Hobbs} G.~B., 
{Edwards} R.~T., {Manchester} R.~N., {Bhat} N.~D.~R., {Sarkissian} J.~M., {Jacoby} B.~A., {Kulkarni} S.~R.,
2008, ApJ, 679, 675

\bibitem[\protect\citeauthoryear {Verbiest, Lorimer, 
\& McLaughlin}{2010}]{vlm10}
{Verbiest} J.~P.~W., {Lorimer} D.~R., {McLaughlin} M.~A., 2010, 
MNRAS, 405, 564

\bibitem[\protect\citeauthoryear {Weisberg, Nice \& Taylor}{2010}]{wnt10}
{Weisberg} J.~M., {Nice} D.~J., {Taylor} J.~H., 2010, ApJ, 722, 1030

\bibitem[\protect\citeauthoryear {Will}{1993}]{wil93}
{Will} C.~M., 1993, 
``Theory and Experiment in Gravitational Physics''
(Cambridge: Cambridge University Press)

\bibitem[\protect\citeauthoryear{Will}{2006}]{will06}
Will, C.~M., 2006, 
``The Confrontation between General Relativity and Experiment'', 
Living Rev. Relativity 9,  (2006),  3. URL: 
http://www.livingreviews.org/lrr-2006-3

\end{thebibliography}
\end{document}